\documentclass[apj,twocolumn,tighten]{openjournal}

\usepackage{lipsum}
\usepackage{newtxtext,newtxmath,enumitem,multirow}

\usepackage{xcolor}
\usepackage{textgreek}
\usepackage[utf8]{inputenc}
\usepackage[english]{babel}
\usepackage{graphicx}	
\usepackage{amsmath}	
\usepackage{amssymb}	
\usepackage{wasysym}    

\usepackage[T1]{fontenc}

\usepackage{hyperref}
\hypersetup{
    colorlinks=true,
    linkcolor=blue,
    citecolor=blue,
    filecolor=magenta,      
    urlcolor=cyan,
    pdfpagemode=FullScreen,
    }

\usepackage{booktabs}
\usepackage{placeins}
\usepackage{capt-of}
\usepackage{csvsimple}
\usepackage{orcidlink}
\usepackage{xspace}

\makeatletter 
  \patchcmd{\NAT@citex}
    {\@citea\NAT@hyper@{%
      \NAT@nmfmt{\NAT@nm}%
      \hyper@natlinkbreak{\NAT@aysep\NAT@spacechar}{\@citeb\@extra@b@citeb}%
      \NAT@date}}
    {\@citea\NAT@nmfmt{\NAT@nm}%
    \NAT@aysep\NAT@spacechar\NAT@hyper@{\NAT@date}}{}{}

  \patchcmd{\NAT@citex}
    {\@citea\NAT@hyper@{%
      \NAT@nmfmt{\NAT@nm}%
      \hyper@natlinkbreak{\NAT@spacechar\NAT@@open\if*#1*\else#1\NAT@spacechar\fi}%
        {\@citeb\@extra@b@citeb}%
      \NAT@date}}
    {\@citea\NAT@nmfmt{\NAT@nm}%
    \NAT@spacechar\NAT@@open\if*#1*\else#1\NAT@spacechar\fi\NAT@hyper@{\NAT@date}}
    {}{}
\makeatother

\newcommand{\lya}{Ly$\alpha$}
\newcommand{\ha}{H$\alpha$}

\newcommand{\oiii}{[O\,\textsc{iii}]}

\newcommand{\nii}{[N\,\textsc{ii}]}

\newcommand{\sii}{[S\,\textsc{ii}]}
\newcommand{\siii}{[S\,\textsc{iii}]}
\newcommand{\oi}{[O\,\textsc{i}]}

\newcommand{\hei}{He\,\textsc{i}}
\newcommand{\almacii}{[C\,\textsc{ii}]}

\newlength{\wdth}

\newcommand{\kms}{km\,s$^{-1}$}
\newcommand{\flux}{erg\,s$^{-1}$\,cm$^{-2}$}
\newcommand{\sfr}{$M_\odot$\,yr$^{-1}$}

\shorttitle{Assembly of a massive galaxy at $z\sim4$}
\shortauthors{Saxena et al.}

\begin{document}
\title{JWST observes the assembly of a massive galaxy at $z\sim4$\vspace{-1.3cm}}

\author{Aayush Saxena$^{1,\star}$ $\orcidlink{0000-0001-5333-9970}$,
Roderik A. Overzier$^{2,3,4}$ $\orcidlink{0000-0002-8214-7617}$,
Catarina Aydar$^{5,6}$ $\orcidlink{0000-0001-5609-2774}$,
Jianwei Lyu$^{7}$ $\orcidlink{0000-0002-6221-1829}$,
George H. Rieke$^{7}$ $\orcidlink{0000-0003-2303-6519}$, \\
Victoria Reynaldi$^{8,9}$ $\orcidlink{0000-0002-6472-6711}$,
Montserrat Villar-Mart\'{i}n$^{10}$,
Krisztina \'{E}va Gab\'{a}nyi$^{11,12,13,14,15}$, \\
Kenneth J. Duncan$^{16} $\orcidlink{0000-0001-6889-8388}$,$
S\'{a}ndor Frey$^{11,13,14}$ $\orcidlink{0000-0003-3079-1889}$,
Andrew Humphrey$^{17,18}$,
George Miley$^{2}$,
Laura Pentericci$^{19}$,
Krisztina Perger$^{13,14}$ $\orcidlink{0000-0002-6044-6069}$,
Huub R\"{o}ttgering$^{2},$
Philip Best$^{15}$,
Sarah E. I. Bosman$^{20,21}$,
Gyorgy Mez\H{o}$^{13,14}$ $\orcidlink{0000-0002-0686-7479}$,
Masafusa Onoue$^{22,23,24}$, \\
Zsolt Paragi$^{25}$ $\orcidlink{0000-0002-5195-335X}$,
Bram Venemans$^{2}$
\vspace{0.2cm}}
\thanks{$^\star$ Email: \href{mailto:aayush.saxena@physics.ox.ac.uk}{aayush.saxena@physics.ox.ac.uk}}

\affiliation{$^{1}$Department of Physics, University of Oxford, 
Denys Wilkinson Building, Keble Road, Oxford OX1 3RH, UK \\
$^{2}$Leiden Observatory, University of Leiden, Niels Bohrweg 2, 2333 CA Leiden, The Netherlands \\
$^{3}$Observat\'orio Nacional/MCTI, Rua General Jos\'e Cristino 77, Rio de Janeiro, RJ 20921-400, Brazil \\
$^{4}$TNO, Oude Waalsdorperweg 63, 2597 AK, Den Haag, The Netherlands \\
$^{5}$Max Planck Institute for Extraterrestrial Physics, Giessenbachstrasse 1, 85748 Garching, Germany \\
$^{6}$Excellence Cluster ORIGINS, Boltzmannstrasse 2, D-85748 Garching, Germany \\
$^{7}$Steward Observatory, University of Arizona, 933 North Cherry Avenue, Tucson, AZ 85721, USA \\
$^{8}$Instituto de Astrof\'isica de La Plata, CONICET-UNLP, Paseo del Bosque, B1900FWA La Plata, Argentina \\
$^{9}$Facultad de Ciencias Astron\'omicas y Geofísicas, Universidad Nacional de La Plata, Argentina \\
$^{10}$Centro de Astrobiolog\'{i}a (CAB), CSIC-INTA, Ctra. de Ajalvir, km 4, 28850 Torrej\'{o}n de Ardoz, Madrid, Spain \\
$^{11}$Department of Astronomy, Institute of Physics and Astronomy, ELTE Eo\"tov\"os Lor\'and University, P\'azm\'any P\'eter S\'et\'any 1/A, H-1117, Budapest, Hungary \\
$^{12}$HUN-REN–ELTE Extragalactic Astrophysics Research Group, ELTE Eo\"tov\"os Lor\'and University, P\'azm\'any P\'eter S\'et\'any 1/A, H-1117, Budapest, Hungary \\
$^{13}$Konkoly Observatory, HUN-REN Research Centre for Astronomy and Earth Sciences, Konkoly Thege Mikl\'os \'ut 15-17, H-1121 Budapest, Hungary \\
$^{14}$CSFK, MTA Centre of Excellence, Konkoly Thege Mikl\'os \'ut 15-17, H-1121 Budapest, Hungary \\
$^{15}$Institute of Astronomy, Faculty of Physics, Astronomy and Informatics,
Nicolaus Copernicus University, Grudzi\c adzka 5, 87-100 Toru\'n, Poland \\
$^{16}$Institute for Astronomy, University of Edinburgh Royal Observatory, Blackford Hill, Edinburgh, EH9 3HJ, UK \\
$^{17}$DTx -- Digital Transformation CoLab, Building 1, Azur\'em Campus, University of Minho, 4800-058 Guimar\~aes, Portugal \\
$^{18}$Instituto de Astrof\'isica e Ci\^encias do Espa\c{c}o, Universidade do Porto, CAUP, Rua das Estrelas, PT4150-762 Porto, Portugal \\
$^{19}$INAF, Osservatorio Astronomico di Roma, via Frascati 33, 00078, Monteporzio Catone, Italy\\
$^{20}$Institute for Theoretical Physics, Heidelberg University, Philosophenweg 12, D-69120, Heidelberg, Germany\\
$^{21}$Max-Planck-Institut f\"ur Astronomie, K\"onigstuhl 17, D-69117, Heidelberg, Germany\\
$^{22}$Kavli IPMU, WPI, The University of Tokyo, 5-1-5 Kashiwanoha,
Kashiwa, Chiba 277-8583, Japan\\
$^{23}$Center for Data-Driven Discovery, Kavli IPMU (WPI), UTIAS, The University of Tokyo, Kashiwa, Chiba 277-8583, Japan\\
$^{24}$Kavli Institute for Astronomy and Astrophysics, Peking University, Beijing 100871, P.R.China\\
$^{25}$Joint Institute for VLBI ERIC, Oude Hoogeveensedijk 4, 7991 PD Dwingeloo, The Netherlands\\
}

\begin{abstract}
We present \textit{JWST} observations of the radio galaxy TGSSJ1530+1049, spectroscopically confirmed at $z=4.0$. NIRCam images and NIRSpec/IFU spectroscopy ($R=2700$) show that TGSSJ1530$+$1049 is part of one of the densest-known structures of continuum and line-emitting objects found at these redshifts. NIRCam images show a number of distinct continuum objects and evidence of interactions traced by diffuse emission, and the NIRSpec IFU cube reveals further strong line emitting regions. We identify six continuum and four additional strong H$\alpha$ emitting sources with weaker or no underlying continuum within the $3\arcsec\times3\arcsec$ IFU field. From spatial alignment with high-resolution radio data and emission line profiles, the radio AGN host galaxy is clearly identified. The bright H$\alpha$ emission (but not the optical components) is distributed remarkably linearly along the radio axis, suggestive of a biconical illumination zone by a central obscured AGN. The emission line kinematics indicate jet-gas interactions on scales of a few kpc. However, due to large relative velocities and presence of underlying continuum, the alignment with the radio structure appears to be, at least partly, caused by a particular configuration of interacting galaxies. At least four objects within a $10\times10$ (projected) kpc$^2$ area which includes the radio source have high stellar masses ($\log(M_\star/M_\odot)>10.3$) and star formation rates in the range $70-163\,M_\odot$\,yr$^{-1}$. Using a stellar mass-based analysis, we predict a total dark matter halo mass of $\approx10^{13}\,M_\odot$. Based on the physical separations and velocity differences between the galaxies, it is expected that these galaxies will merge to form a massive galaxy within a few Gyr. The system qualitatively resembles the forming brightest cluster galaxies in cosmological simulations that form early through a rapid succession of mergers.
\end{abstract}


\maketitle

\section{Introduction} 
\label{sec:intro}

High-redshift ($z>2$) radio galaxies (HzRGs) are the host galaxies of powerful radio synchrotron sources in the early Universe \citep{mccarthy93}. Their radio emission is associated with large jets of relativistic plasma that are driven by activity of a central supermassive black hole \citep[SMBH;][]{rees84,blandford19}. Unlike quasars at high redshift, in which the emission from the active galactic nucleus (AGN) easily outshines the stellar continuum emission from the host galaxy \citep{peng06,ding23,onorato25}, HzRGs tend to offer a clear view of their host galaxy including dust, nebular gas, and stellar populations \citep[e.g.,][]{miley08}. This makes them useful for understanding the relation between the activity of SMBHs and the formation and evolution of the host galaxies.

However, especially at high redshifts, it is often still not straightforward to disentangle AGN-related processes from typical processes that occur in the host galaxy. Relativistic radio jets can trigger shocks and drive outflows that interact with neutral or ionized gas and dust, with direct or indirect consequences for the process of star formation \citep{begelman89, Heckman24}. At the same time, star formation or merging activity in the (predominantly) massive host galaxies can further interact with the gas or alter the spatial distributions of dust, metals, and stars. Due to jet-gas interactions, the morphology, kinematics and ionization properties of the emission line gas are often peculiar, thus requiring high spatial resolutions to interpret all the different features and components \citep[e.g.,][]{baum92,villar-martin97,best00,nesvadba17a,saxena24}.

HzRGs are known to be associated with the massive ends of both SMBH and host galaxy populations \citep[see review by][]{miley08}. They have also been shown to probe relatively dense galactic environments, including protoclusters \citep{venemans07,wylezalek13,hatch14,overzier16}. This makes HzRGs beacons of protocluster environments at high redshift, further motivating the search for new HzRGs and their follow-up studies. The protocluster environments in which some HzRGs have been found also support their interpretation as progenitors of very massive galaxies, such as the massive elliptical galaxies observed at the centers of galaxy clusters \citep[e.g.][]{west94,overzier03,hatch09,emonts16,retana-montenegro17,Hardcastle2019,sabater19,magliocchetti22,poitevineau23,saxena24}. Therefore, the complex nature of HzRGs and their importance as cosmological probes motivated several programs in the first cycle of observations with \textit{JWST}. Our program (GO 1964) targeted two HzRGs at $z>4$ \citep[][this paper]{roy24,saxena24}.

In this article, we present \emph{JWST}/NIRSpec IFU observations of the HzRG TGSS J1530+1049 \citep[``TGSSJ1530'';][]{saxena18a}. The radio source (at J2000 coordinates 15:30:49.9, +10:49:31.1) was selected as a candidate high redshift object from the 150 MHz TGSS Alternative Data Release 1 \citep{intema2017} survey on the basis of its ultra-steep radio spectral index ($f_\nu\propto\nu^{\alpha}$ with $\alpha^{{150 \mathrm{MHz}}}_{{1.4 \mathrm{GHz}}} = -1.4$) and compact morphology via follow-up VLA imaging at 1.4 GHz. In a companion paper, new radio observations at milliarcsecond (mas) scale angular resolution with the European VLBI Network (EVN), and at $\sim$100 mas-scale resolution with the enhanced Multi-Element Remotely Linked Interferometer Network (e-MERLIN), are presented \citep{gabanyi25}. The new radio data confirm that TGSSJ1530 is a relatively small ($\sim$5.4 kpc projected separation between its hotspots) and symmetric radio source. 

Existing optical imaging from surveys such as SDSS and Pan-STARRS showed no counterpart to the radio source. Follow-up deep near infrared imaging revealed a faint $K$-band detection consistent with the position of the radio source, which using the well-established relationship between the $K$-band magnitude and redshift of radio galaxies pointed towards a likely high redshift nature \citep{saxena18a, saxena18b}. A redshift was determined based on a single faint emission line detected in a Gemini/GMOS-N spectrum. Due to a slight blue asymmetry, lack of continuum emission and expectation of a high redshift of the ultra-steep spectrum radio source, this emission line was interpreted as \lya\ at $z=5.72$ \citep{saxena18a, saxena19}. 

However, our new observations unambiguously reveal that its true redshift is $z\approx4.0$, with no emission line seen that would be expected from a redshift of $z=5.72$. This suggests that either the emission line initially interpreted as \lya\ was spurious, which is unlikely given that the emission line was independently detected in three separate frames from GMOS observations \citep{saxena18a}, or that a background \lya\ source exists outside of the NIRSpec IFU field. However, we note that the emission line interpreted as \lya\ by \citet{saxena18a} is instead consistent with He\,\textsc{ii}\,$\lambda1640$ at a redshift of $z\approx3.98$, aligning with the secure redshifts of the continuum components seen in the TGSS1530 system thanks to high resolution NIRCam imaging and NIRSpec spectroscopy, as we will show in this work. We defer the detailed investigation of the initially reported rest-frame UV line to a future paper. 

Nonetheless, the \textit{JWST} data presented here show a highly peculiar system comprising multiple distinct continuum and line-emitting sources around the radio source. As we show in this paper, the galactic environment surrounding the TGSSJ1530 radio source sheds light on the formation of a very massive galaxy in the early Universe that includes at least one active SMBH.  

In Section \ref{sec:data} we present our \textit{JWST} observations, ancillary data, and details on the processing. In Section \ref{sec:results} we present the main measurements based on the 1D spectra and the 2D line maps extracted from the NIRSpec/IFU data. In Section \ref{sec:massive} we use observations from both NIRSpec IFU and NIRCam to provide an explanation for the peculiar structure of TGSSJ1530 in the context of a massive forming galaxy. In Section \ref{sec:conclusions} we summarize our findings.

Throughout this paper we assume a flat $\Lambda$CDM cosmology as determined by the \citet[][$H_0 = 67.7$ km s$^{-1}$ Mpc$^{-1}$, $\Omega_m = 0.31$]{planck20}, giving a cosmic age of 1.5 Gyr and a spatial scale of $7.1$ proper kpc\,arcsec$^{-1}$ at $z=4.0$. 

\section{Data} 
\label{sec:data}

\subsection{JWST/NIRSpec IFU observations}

TGSSJ1530 was observed with the \textit{JWST}/NIRspec in IFU mode on July 14-15 (UT), 2023 (Program GO1964; Co-PIs R. Overzier and A. Saxena). The high resolution ($R\sim2700$) grating/disperser F290LP/G395H was used. Nodding in-scene was performed to maximize the on-source exposure time and determine the background locally and in real-time. We observed 48 groups per integration and a 4-point dither with 2 integrations at each point for a total exposure time of 28 ksec.

The data reduction procedure used was identical to that described in \citet{saxena24}, which presented observations of another radio galaxy at $z=4$ from the same program. For details about data reduction, we refer the readers to \citet{saxena24} and present a brief overview of the procedure here. 

The Stage 1 products downloaded from the Mikulski Archive for Space Telescopes (MAST) Portal\footnote{\url{https://mast.stsci.edu/portal/}} were gain-corrected, flagged for `snowballs' and jumps, bias corrected and dark current corrected. Stage 2 of the pipeline was then run on these zero-corrected frames, and the processing steps included world coordinate system (WCS) correction, background subtraction (no dedicated background observation was taken for the science cube), flat-fielding, pathloss correction, distortion correction and photometric calibration. After Stage 2 processing, a few iterations of cosmic ray/outlier rejection were performed using the \texttt{lacosmic} algorithm\footnote{\url{https://lacosmic.readthedocs.io/en/stable/}} \citep{lacosmic} on all individual 2D frames. Stage 3 processing was then performed, which included the 3D data cube building step, where we used the ``Shephard's method'' weighting option in the pipeline. The final 3D cube was visually inspected and any remaining bright pixels were manually flagged. 

\subsection{JWST/NIRCam observations}
\begin{figure*}
    \centering
    \includegraphics[width=\textwidth]{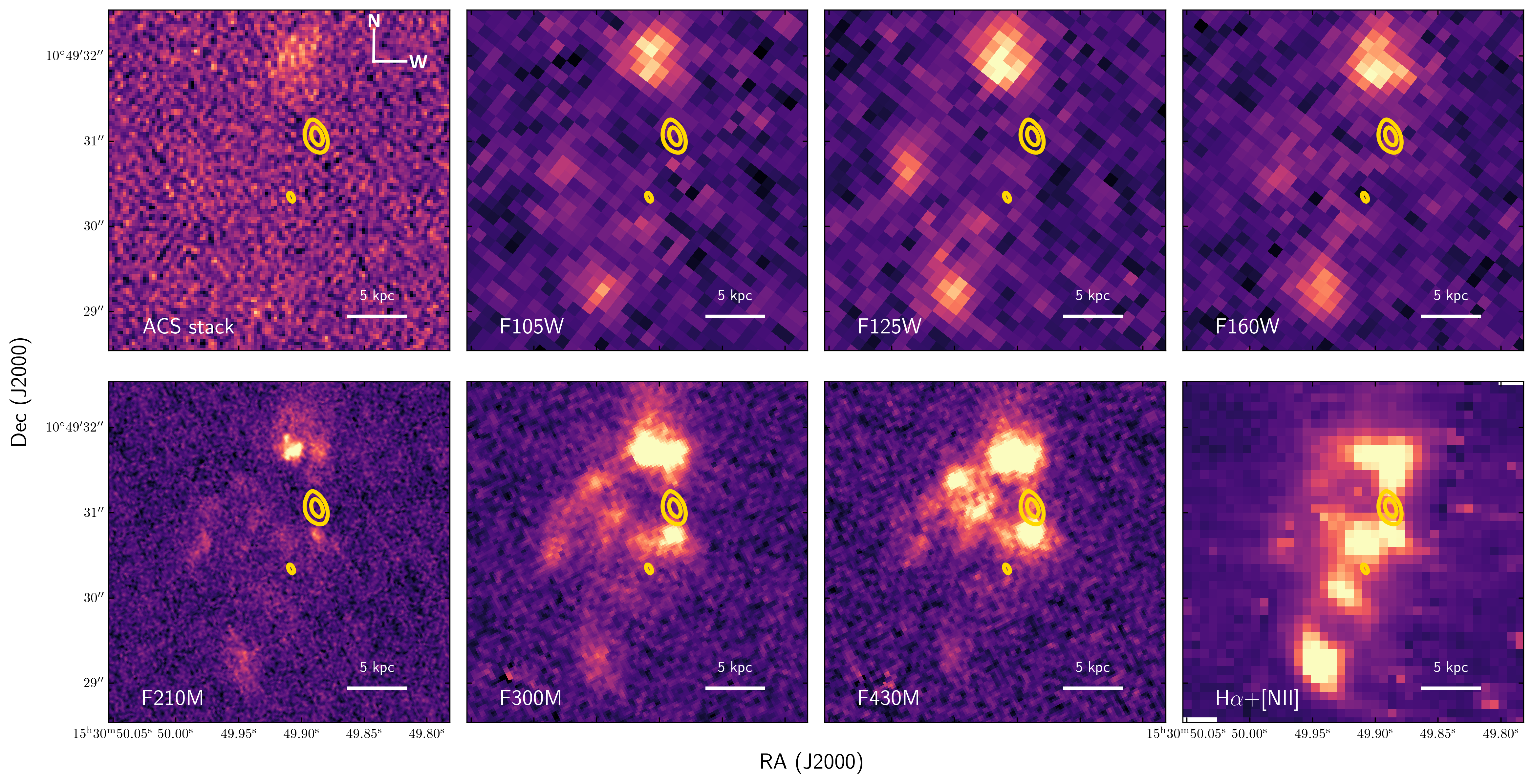}
    \caption{Top-left to bottom-right: Median stacked HST/ACS images (F606W, F775W, F850LP), HST/WFC3 images (F105W, F125W, F160W), JWST/NIRCam images (F210M, F300M, F430M) and the JWST/NIRSpec IFU image of \ha+\nii. In all panels, the radio VLBI contours at 1.4\,GHz from \citet{gabanyi25} are shown. The NIRCam images show a highly complex morphology including multiple clumps and diffuse (continuum) emission that may be indicative of mergers and interactions. The \ha+\nii\ image on the other hand shows a remarkable linear configuration of strong line-emitting sources that tend to be faint or absent in the continuum images. The variation in brightness over the larger wavelength range shows that the SEDs of the individual sources in this complex system vary substantially. The diameter of the maps correspond to a projected physical size of about 21 kpc at $z=4$.}
    \label{fig:images}
\end{figure*}

NIRCam imaging observations covering the field toward TGSSJ1530 were carried out on July 17 (UT), 2023 (Program GTO1205; PI G. Rieke). Two sets of observations were obtained, one using the F210M and F300M filters in the short and long wavelength channels, and another one using the F210M and F430M filters. A 4-point dither pattern was used with two integrations per exposure and 10 groups per integration in the MEDIUM8 readout pattern. The total exposure times were 2.6\,ksec in F210M, and 1.3 ksec each in F300M and F430M. We processed the NIRCam data with the \textit{JWST} pipeline v1.9.6 \citep{Bushouse2023} and Calibration Reference Data System (CRDS) \texttt{jwst\_1084.pmap}. Besides the standard reduction steps, we added a custom step in Stage 2 of the pipeline to characterize and subtract the $1/f$ noise from each frame. The final mosaic images were resampled to smaller pixel scales (0.\arcsec 0147 pixel$^{-1}$ in the SW filters; 0.\arcsec03 pixel$^{-1}$ in the LW filters) using the drizzling algorithm in Stage 3.

Owing to the small field-of-view of the NIRSpec/IFU cubes, we used the NIRCam imaging to astrometrically register the NIRSpec data. We convolved the IFU cube with the NIRCam F210M, F300M and F430M medium band filter transmission curves to generate  synthetic NIRCam images. These were aligned to the actual NIRCam images by minimizing residuals in a normalised difference image. The comparison resulted in a correction of $\approx0.1$\,arcsecond along both the Right Ascension and Declination axes of the final NIRSpec IFU cube.

\subsection{HST observations}

\textit{HST} imaging observations with the ACS and the WFC3 were obtained between March 2022 and February 2023 as part of program GO16693 (PI: R. Overzier). The ACS filters used were F606W (6.3 ksec), F775W (8.3 ksec) and F850LP (10 ksec), and the WFC3 filters used were F105W, F125W and F160W (4.2 ksec each). The images were pipeline reduced. A stack of the ACS images revealed little useful information (see Figure \ref{fig:images}). The WFC3 images show three components and diffuse emission surrounding the general location of the radio source, but no object was clearly identified with this source. As we will show below, the NIRCam images and NIRSpec IFU data were essential to make the correct identifications. 

\subsection{Identification of continuum and line-emitting components}
\begin{figure*}
    \centering
    \includegraphics[width=\linewidth]{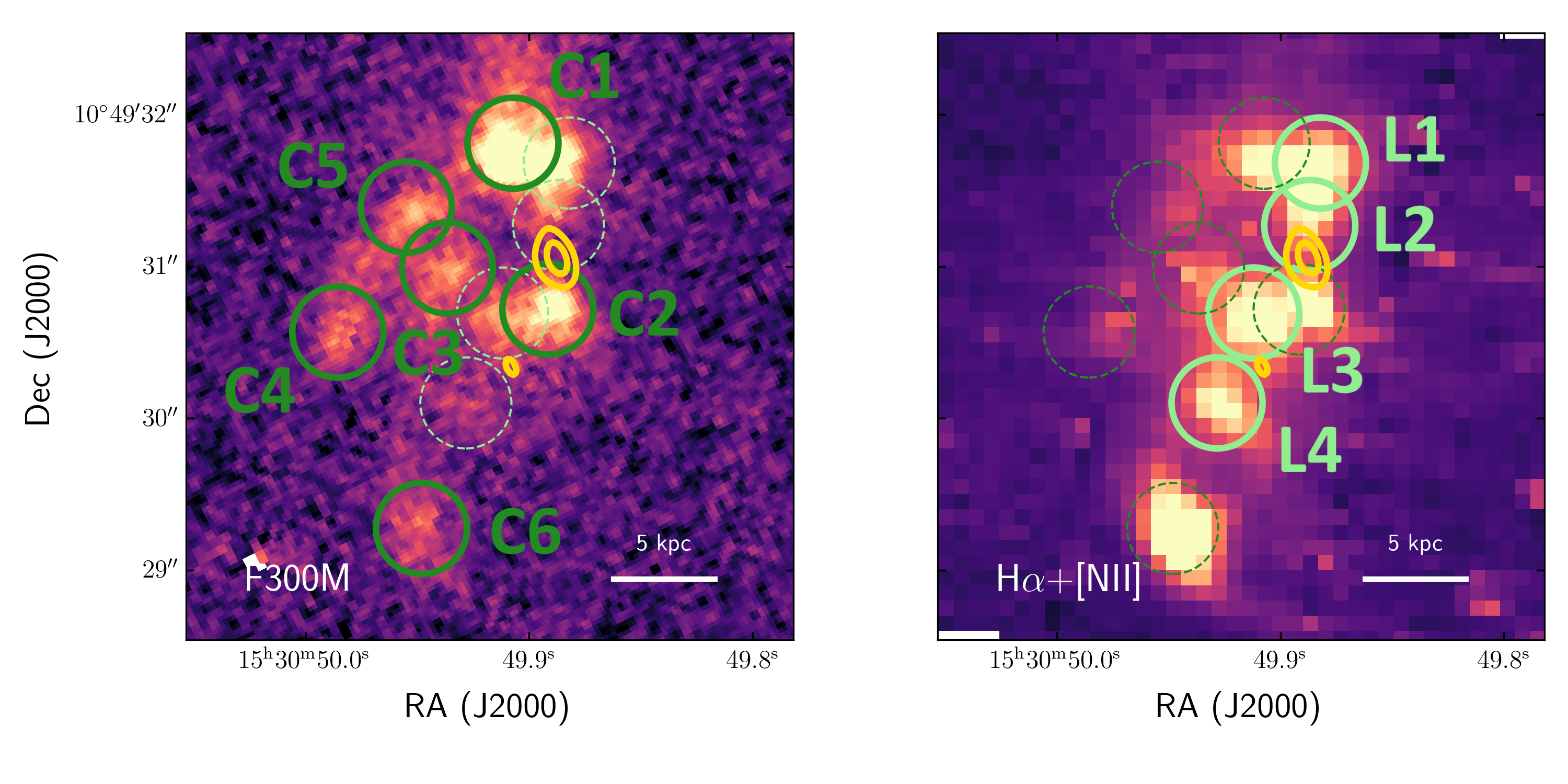}
    \caption{\ha+\nii\ 2D image (left) and F300M map (right) of the system, with  apertures for the spectral extraction indicated (circles). The extraction is performed for six clear continuum sources (C1-C6) identified from the NIRCam image, and four further emission line regions (L1-L4) that show up prominently in the \ha+\nii\ map but appear absent or less prominent in the NIRCam images. Note that one of the brightest line-emitting clumps, at the Southern edge of the image, corresponds to a continuum source that had already been labeled C6. The width of the maps corresponds to a projected physical size of about 21 kpc at $z=4$.}
    \label{fig:nc_ns_apertures}
\end{figure*}

In Figure \ref{fig:images} we show the gallery of \emph{HST}/ACS, \emph{HST}/WFC3 and \textit{JWST}/NIRCam images, along with a continuum subtracted emission line map centered around \ha+\nii\ emission from NIRSpec/IFU (bottom-right panel). The images are largely free of (strong) emission lines (H$\gamma$ and H$\delta$ fall in F210M), allowing the imaging data to be used for inferring optical continuum morphologies. The WFC3, NIRCam and \ha+\nii\ images show multiple components that have been labeled according to the scheme presented in Figure \ref{fig:nc_ns_apertures}, where we show the continuum-subtracted \ha+\nii\ image (left panel) and the F300M continuum image (right panel). 

We first identified the main concentrations of continuum emission in the F300M image, which are labeled C1 to C6. Next, we identified four strong line emitting sources that did not have a clear, well-defined counterpart in the continuum, which are labeled L1 to L4. In total, there are 10 sources of interest. A general trend is that the strongest continuum sources are often not the strongest line sources, and vice versa. We note, in particular, that one of the brightest line-emitting clumps, at the Southern edge of the image, coincides with an extended continuum source that had already been labeled C6. Although different choices could have been made in this labeling scheme, the choices we made do not affect the main results of this paper. We placed circular apertures of radius $0\farcs2$ around the components to extract the 1D spectra.

\begin{figure*}
    \centering
    \includegraphics[width=\linewidth]{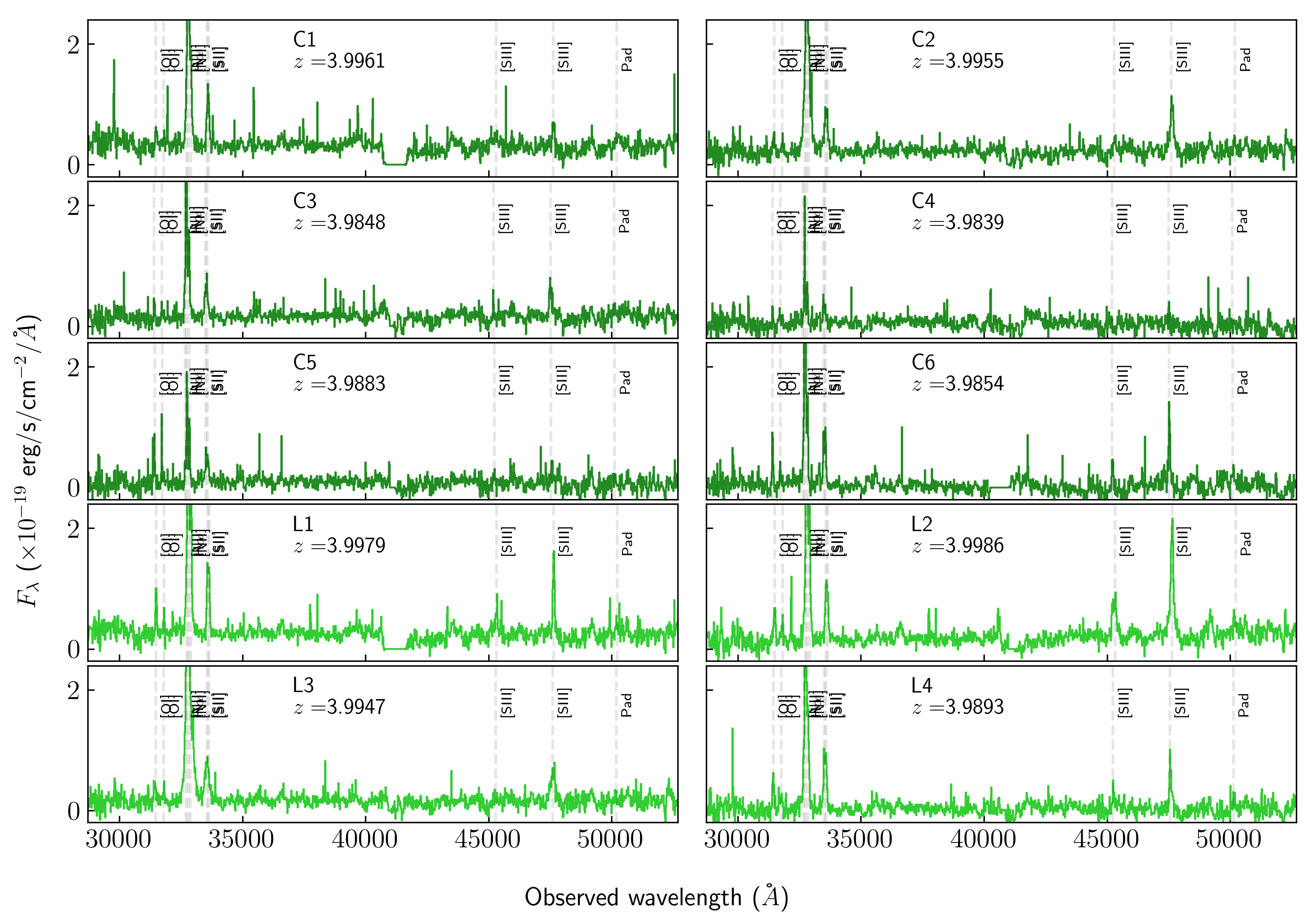}
    \caption{Extracted 1D spectra from continuum sources identified in NIRCam imaging (C1--C6) and line emitting sources identified from NIRSpec (L1--L4), with the locations of prominent emission lines marked. The continuum source C2 is the presumed radio AGN host based on its co-location with the radio emission seen from this system as well as the presence of a broad component in its \ha+\nii\ complex.}
    \label{fig:extraction1D}
\end{figure*}

In Figure \ref{fig:extraction1D} we show 1D spectra extracted from the circular apertures of Figure \ref{fig:nc_ns_apertures}. Across all continuum and line emitting regions, prominent emission lines of the \oi\,$\lambda\lambda 6300,6364$ doublet, \ha, the \nii\,$\lambda\lambda 6548,6583$ doublet, the  \sii\,$\lambda\lambda 6716,6731$ doublet, \hei\,$\lambda 7281$, \siii\,$\lambda 9531$ and (weak) Pa$\mathrm{\delta}$ are seen (marked by the vertical dashed lines in Figure \ref{fig:extraction1D}). Thanks to the detection of multiple strong emission lines, accurate spectroscopic redshifts could be obtained for each component. Redshifts were determined by identifying the peaks of the emission lines and taking the median redshift recorded from each line. Spectroscopic redshift measurements for all the components are given in Table \ref{tab:line_measurements}.

\section{Results}
\label{sec:results}
\subsection{A complex morphology - evidence of merger?}

The WFC3 images, probing the spectral region from the near-UV to just shortwards of the Balmer break, are dominated by the Northern (C1) and Southern (C6) clumps, with C4 also visible. The NIRCam F300M and F430M images that probe the spectrum well beyond the Balmer break show a complex region of prominent clumps and diffuse emission within 1\arcsec\ of the C1 component (C2-C5). Owing to the lack of depth in the ACS images, only C1 can be clearly seen. 

We note that component C1 becomes more prominent in the F300M and F430M images, additionally showing 4-5 clumps and further diffuse emission. Component C6 has the brightest \ha\ emission. It is the second most prominent source in the WFC3 images probing the rest-UV, and appears as a faint but extended continuum source in the NIRCam images. Clumps C2-C5 are all prominent at NIRCam wavelengths, but faint at the shorter WFC3 wavelengths. 

In F300M, which is the deepest NIRCam image, fainter clumps and diffuse emission, especially a bridge connecting C4 and C5 are suggestive of spiral or tidal structure. The general picture that emerges from the morphologies of continuum and diffuse components is that of a complex, interacting system at $z\approx4.0$. The presence of bright and compact radio emission further confirms the presence of at least one powerful AGN in this complex system, which is most likely located in or near component C2 (see next subsection). In comparison with the continuum morphology, the overall morphology in the \ha\ map is strikingly linear over an extent of about 2\arcsec\ or 14 kpc at $z=4.0$. This morphology is reminiscent of the well-known radio-optical alignment effect in HzRGs, and will also be addressed below.

\subsection{The radio AGN and alignment of the \ha\ emission} 

\begin{figure}
    \centering
    \includegraphics[width=\linewidth]{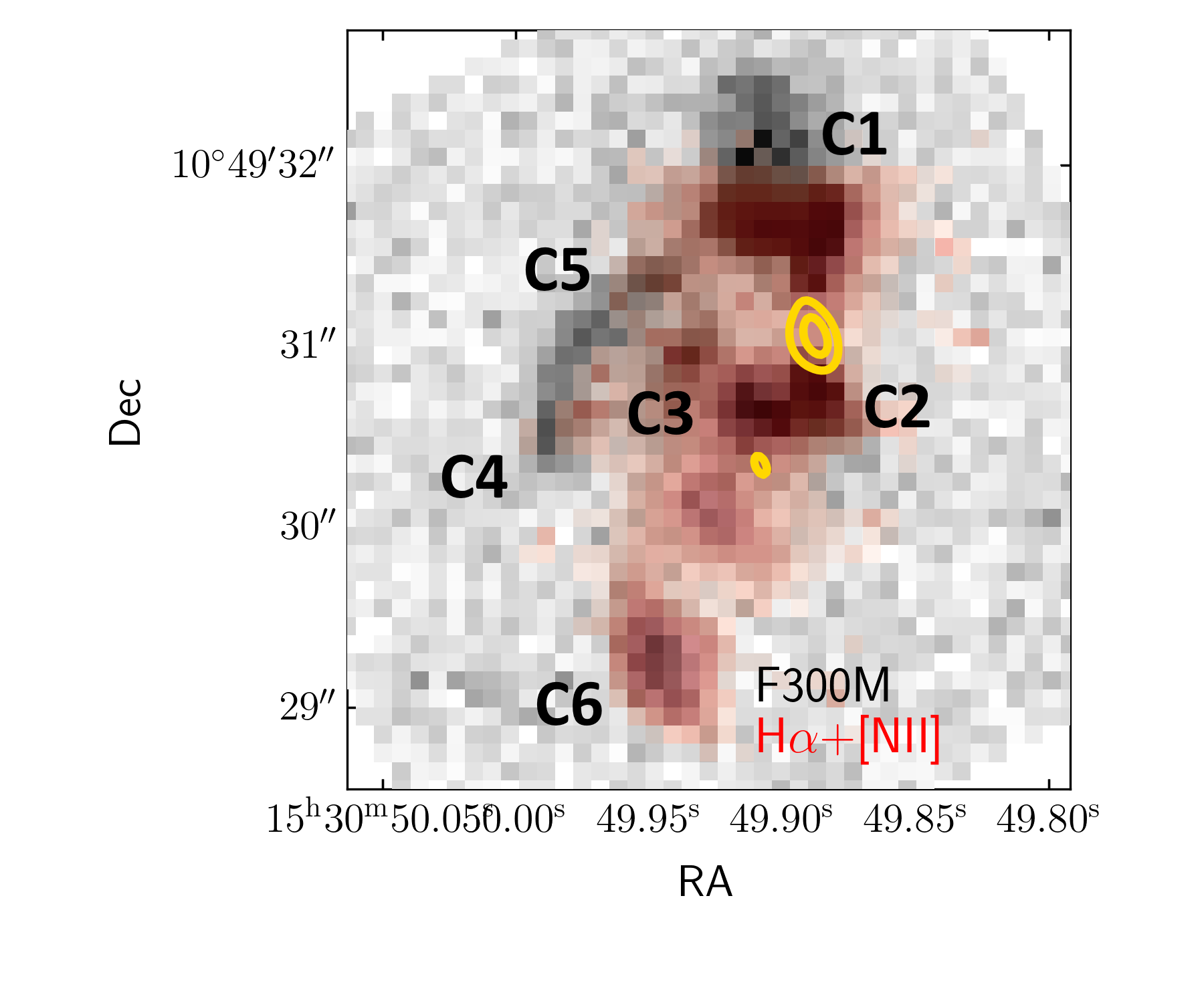}
    \caption{Composite image showing NIRCam F300M in the background, with \ha+\nii\ map overlaid, along with the radio VLBI e-MERLIN L-band contours from \citet{Gabanyi2018,gabanyi25}. This composite highlights the different morphologies seen in the emission line and continuum maps. The radio contours appear to originate from the continuum source C2, which we propose to be the radio AGN host.}
    \label{fig:composite_radio}
\end{figure}

In Figure \ref{fig:composite_radio} we show a composite image based on the F300M continuum flux (background image, in greyscale), the \ha+\nii\ emission (overlaid in red), and the VLBI e-MERLIN/L-band image (yellow contours) from \citet{gabanyi25}. Based on the astrometric solutions for the VLBI data with uncertainties $<10$\,milliarcseconds, the two hotspots identified in the radio image lie on either side of the region containing the second-brightest continuum source, C2. The northern radio hotspot lies in between C2 and L2, while the southern radio hotspot appears to lie in between L3 and L4 (in projection). 

The radio jets, which are not seen in the VLBI data, are expected to lie along the general orientation that is defined by a line connecting the two hotspots observed. This is the same general orientation along which the strong emission line components (but not the continuum components) are found. Using high resolution radio VLBI observations, \citet{gabanyi25} show that both radio hotspots have extremely steep radio spectral indices ($-2.1 \pm 0.1$ for the northern lobe and $-1.7 \pm 0.5$ for the southern lobe), making it highly unlikely that one of these hotspots could be tracing the `core' of the radio galaxy, which typically has a flatter spectral index driven by synchrotron self-absorption of the high-energy jets. 

Alignment of radio structures with primarily the rest-frame UV continuum emission is a well-known effect in HzRGs \citep[e.g.][]{baum89,mccarthy93,best00,tremblay09}. However, the alignment in TGSSJ1530 is clearly restricted to the morphology of the emission line gas. Alignment between the radio jet axis and the emission line gas could arise from the narrow, biconical ionizing radiation that illuminates the gas on either side of the AGN obscuring torus \citep[e.g.][]{tadhunter89,mccarthy93,villar-martin03,humphrey06,miley08,wang24}. Another source of alignment is the interaction between the radio jets and the ISM. In such a scenario, any gas that is already present or driven out by the jets along the radio axis is photoionized or shock-heated \citep[e.g.][]{best00,humphrey06,nesvadba17a,nesvadba17b,saxena24,roy24,wang24}. A third known effect that can lead to radio-optical alignment is the jet-induced star formation. In that case, star formation is triggered along an expanding radio jet, giving rise to young stars and ionized gas \citep[e.g.][]{dey97,gurvits97,bicknell00,salome17,nesvadba20,walter25}.  

In another study that includes the same object \citep{roy25}, it is concluded that the properties of the aligned line emission in TGSSJ1530 is evidence of a $\sim$20 kpc scale galactic outflow and jet-mode feedback. However, that study did not make use of the NIRCam images studied in this paper, and therefore lacked the additional context provided by the optical continuum morphologies across the system. As shown in Figure \ref{fig:composite_radio}, the alignment in TGSSJ1530 is restricted to that between the radio and emission line structures, and not the continuum. Another interesting feature is that the radio-aligned emission line clumps extend well beyond the radio hotspots, making any explanation that depends solely on interaction between the (current) jet activity and the gas unlikely. Also, all of the main emission line clumps are of roughly equal brightness, whereas in the case of a classical ionization cone, one would expect a radial fall-off in the line fluxes with distance from the AGN nucleus. The brightest line emitting region is also an (extended) continuum source (C6), and it lies at the extreme Southern end of the ``line". It is thus unlikely that this is the result of (outflowing) gas that is being photoionized by the central AGN. 

At the extreme Northern end, the line emission originates from near the brightest continuum source in the redder NIRCam bands (C1). At least the continuum emission is difficult to explain in the case of pure photoionization or jet-induced star formation scenarios, as C1 is one of the most massive components (see Sect. \ref{sec:seds}). Both C1 and C6 also have a systematic velocity that is offset from the velocity of C2, by about $+100$ and $-700$\,\kms, respectively (see Sect. \ref{sec:kinematics}). It is thus more likely that these are separate galaxies involved in a merger. If we were to exclude C6, and possibly C1, the alignment of the emission line gas along the radio axis would be significantly less dramatic. However, clumps L2, L3, L4 that lie closest to the central AGN and are aligned with the visible radio emission may very well be explained by central emission along an ionization cone, perhaps with a contribution from jet-gas interactions in the very inner part. 

However, we note here that the present high-resolution radio observations are relatively shallow and may be resolving out any fainter and more diffuse emission in this system. For example, deeper radio follow-up observations of a radio-quiet quasar at $z=0.1$ revealed extended radio emission out to $\sim26$\,kpc from the radio AGN, spatially coinciding with the bulk of line emission seen away from the AGN \citep{villar2017}. If similar diffuse, extended radio emission were to exist on a similar scale in TGSSJ1530, our conclusions about the dominant source of ionization at these scales would have to be revisited. However, we point out that source C6, which is a very strong line emitting source and lies well beyond the visible radio emission, has the clear characteristics of a separate galaxy. 

In summary, the presence of strong continuum sources, the large extent of the line emission well beyond the VLBI radio structure, and the presence of diffuse continuum and line emission surrounding the brighter sources suggests that the morphology of TGSSJ1530 is best explained by a system of merging sub-galactic clumps or galaxies. One of these components hosts the radio AGN, which may further ionize (and interact with) gas in its surroundings along a preferred axis. 

\subsection{Emission line profiles tracing a complex system}

\begin{figure*}
    \centering
    \includegraphics[width=0.24\linewidth]{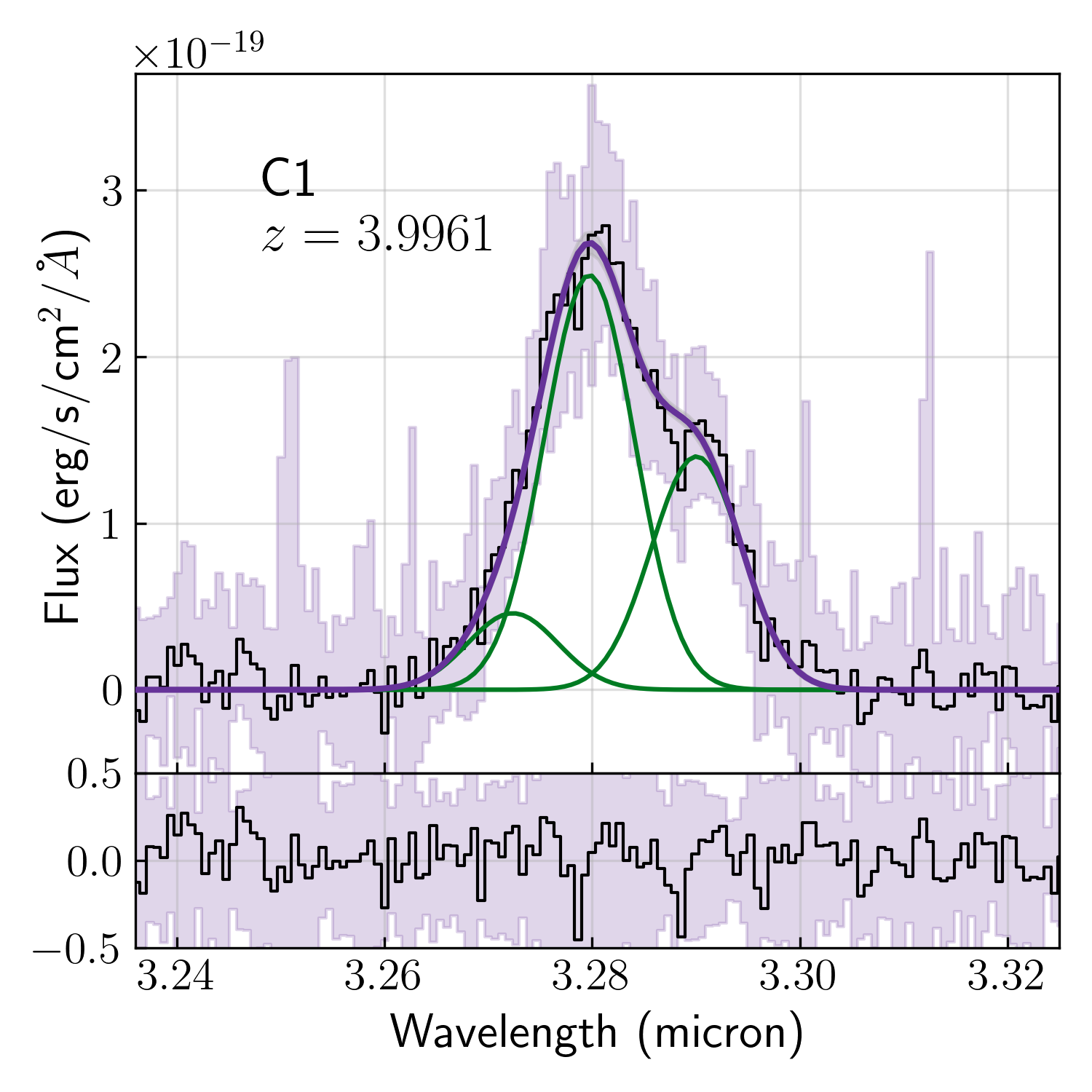}
    \includegraphics[width=0.24\linewidth]{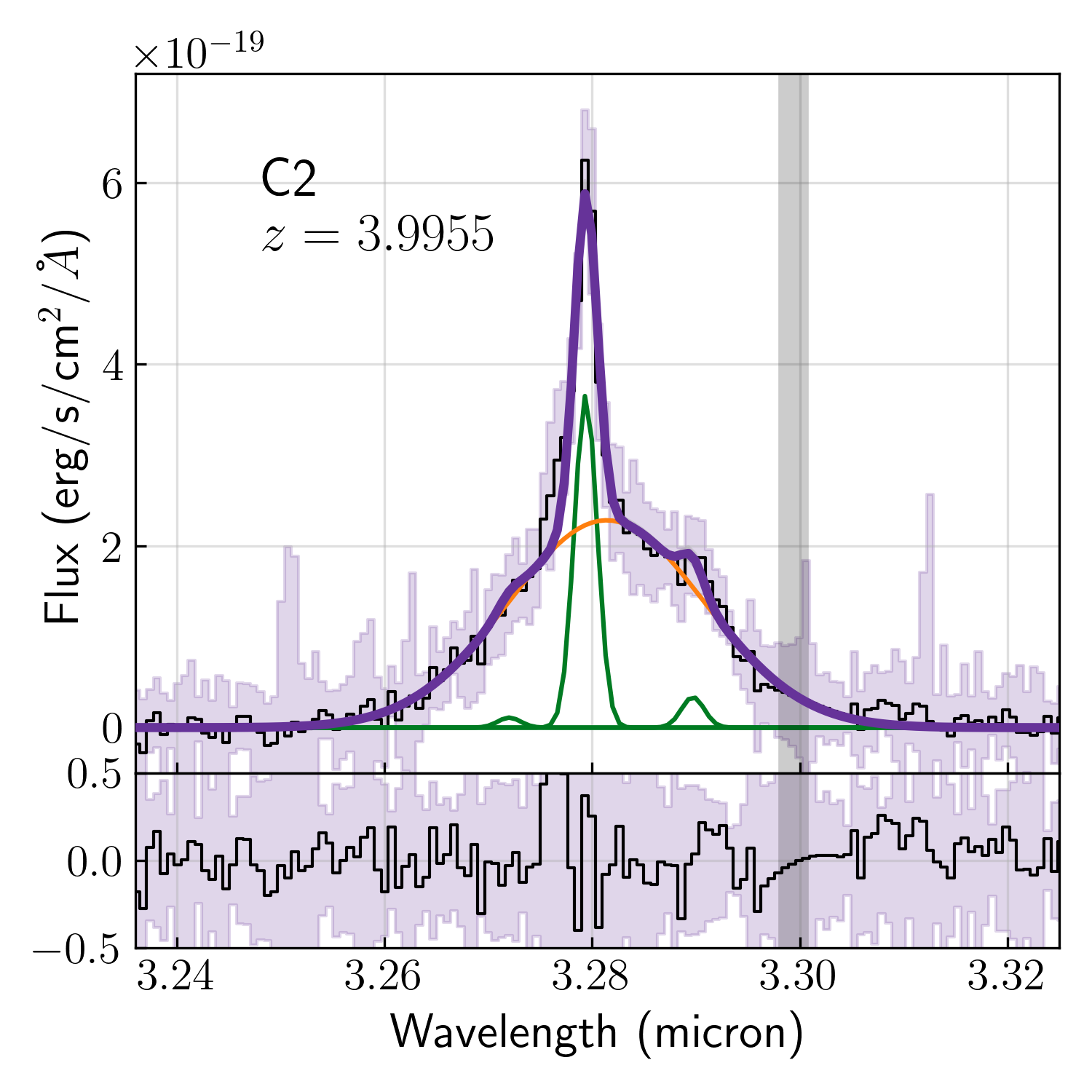}
    \includegraphics[width=0.24\linewidth]{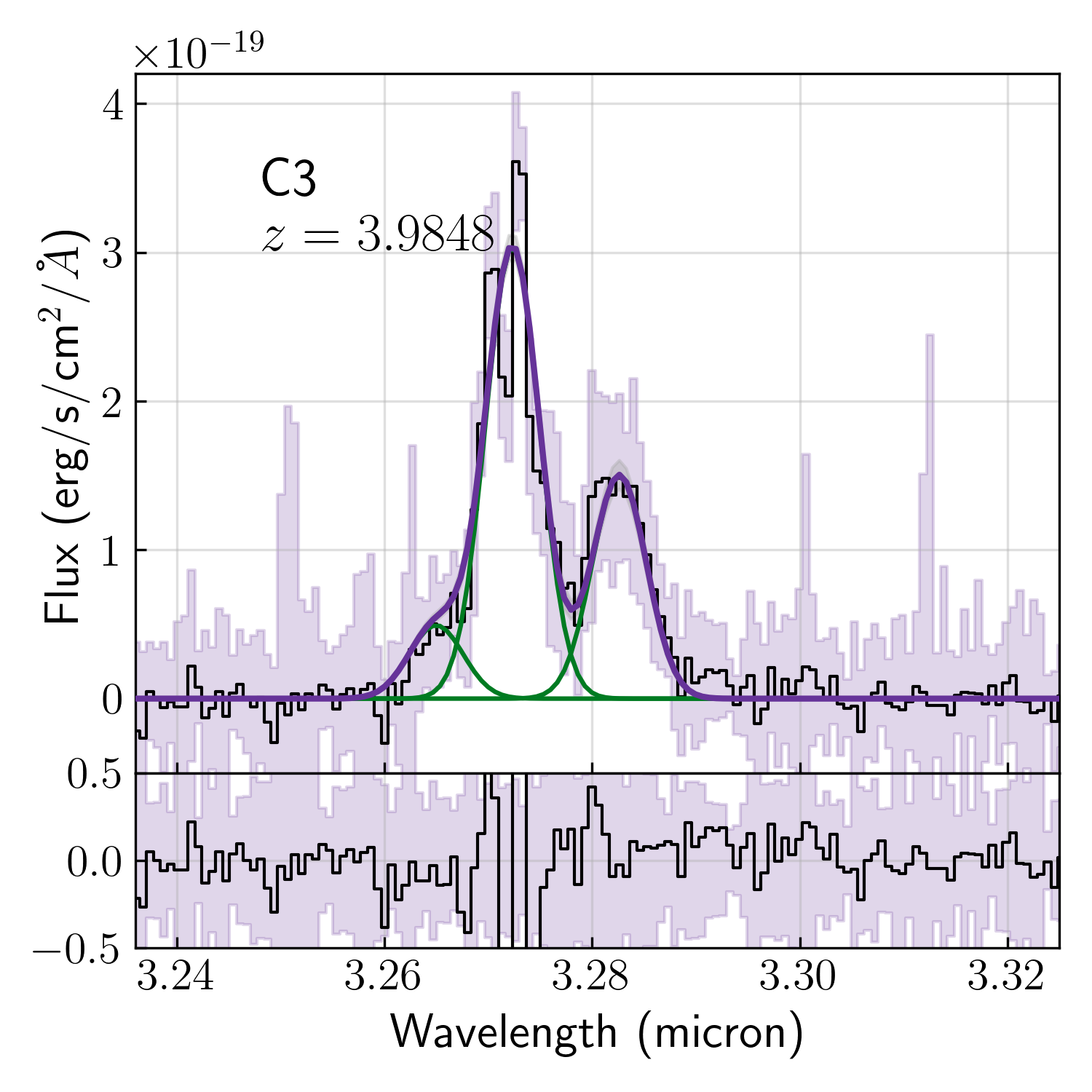}
    \includegraphics[width=0.24\linewidth]{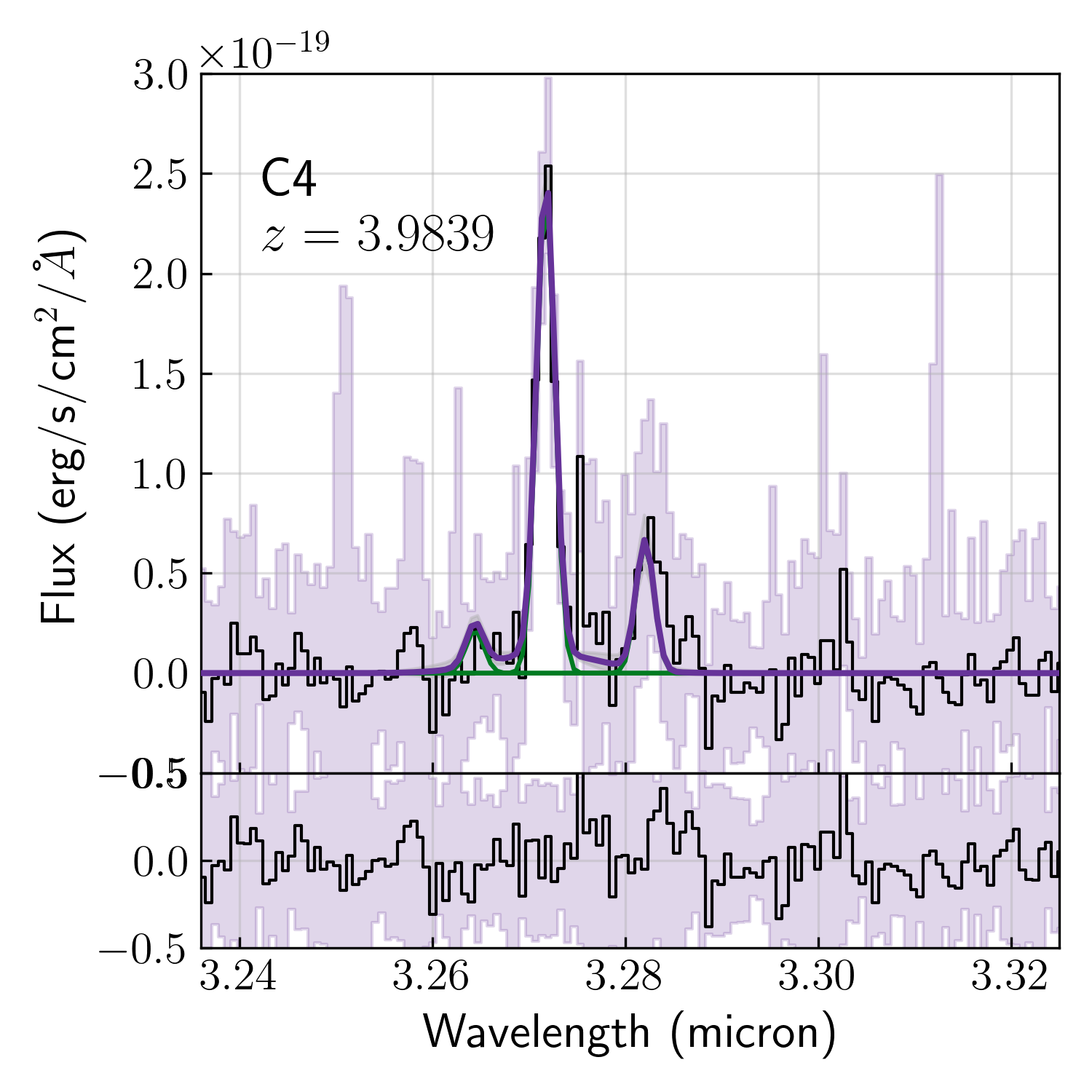}
    
    \includegraphics[width=0.24\linewidth]{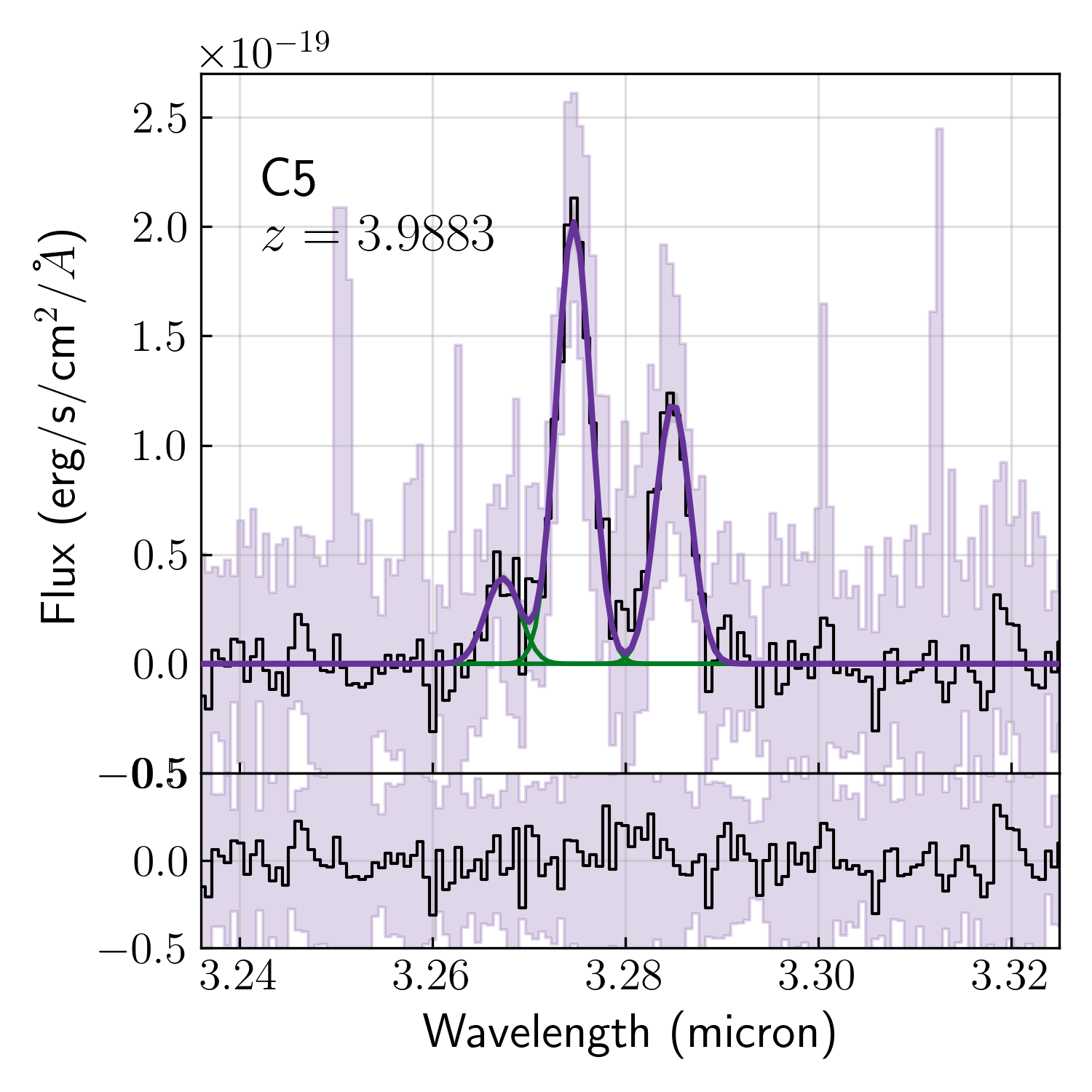}
    \includegraphics[width=0.24\linewidth]{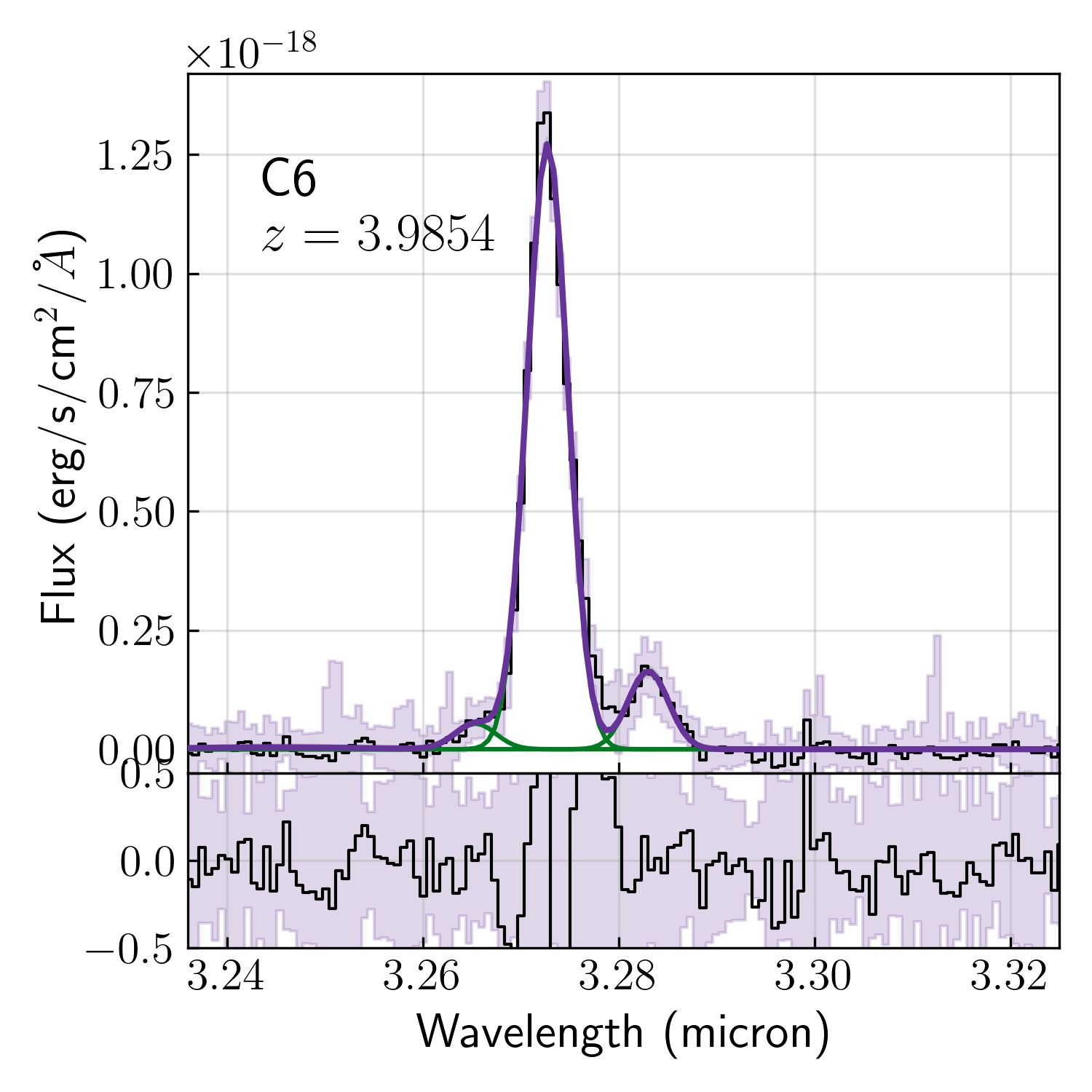}    
    
    \includegraphics[width=0.24\linewidth]{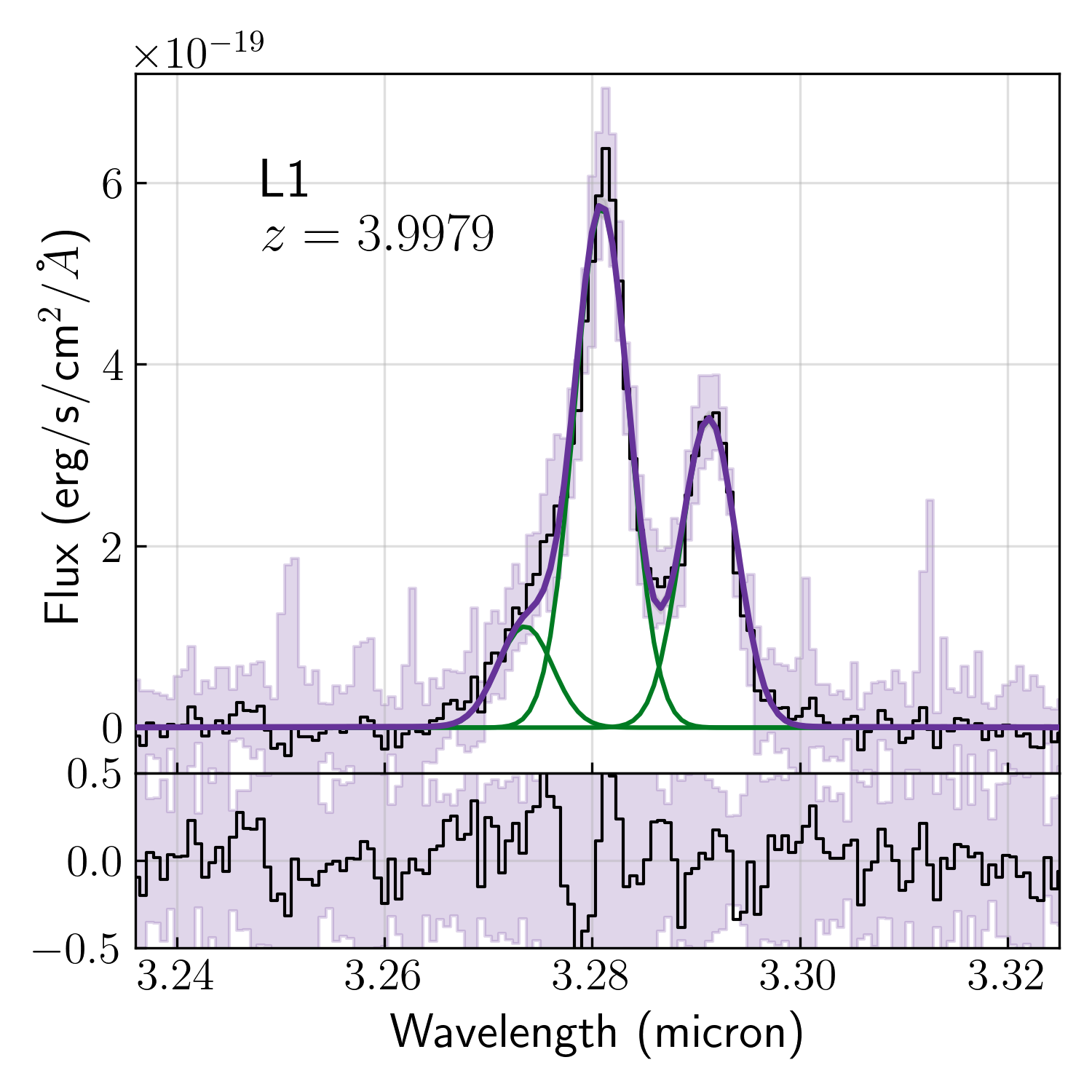}
    \includegraphics[width=0.24\linewidth]{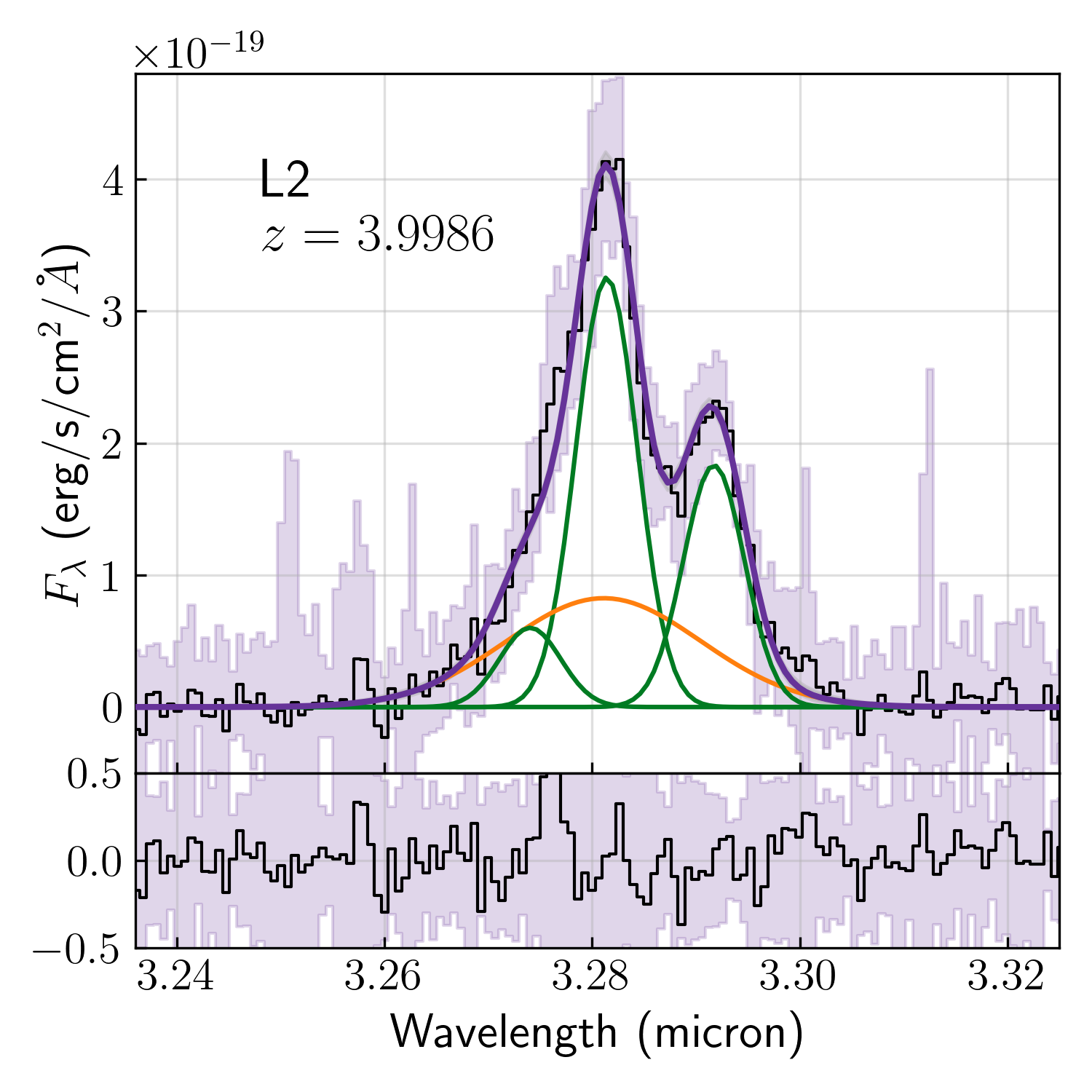}
    \includegraphics[width=0.24\linewidth]{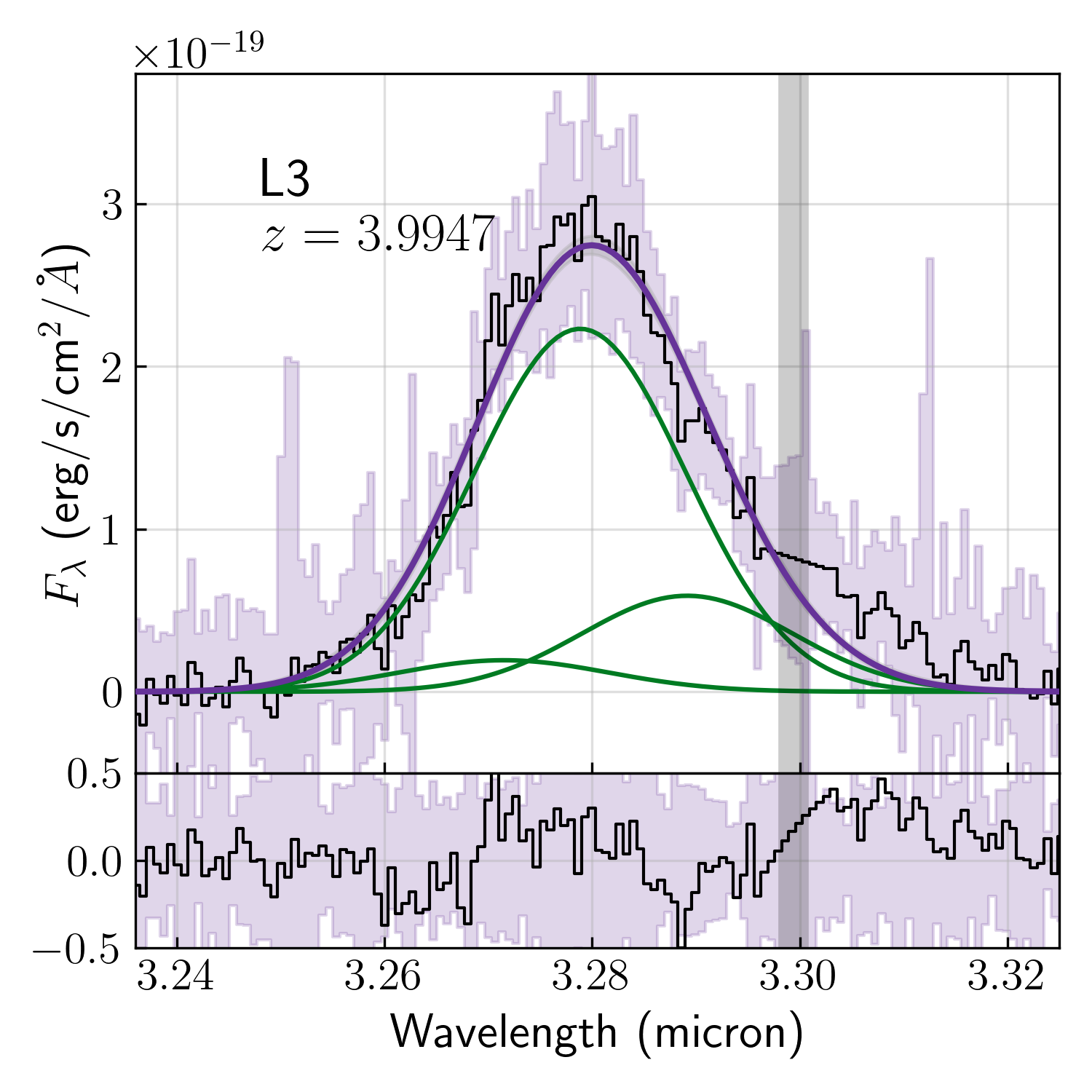}
    \includegraphics[width=0.24\linewidth]{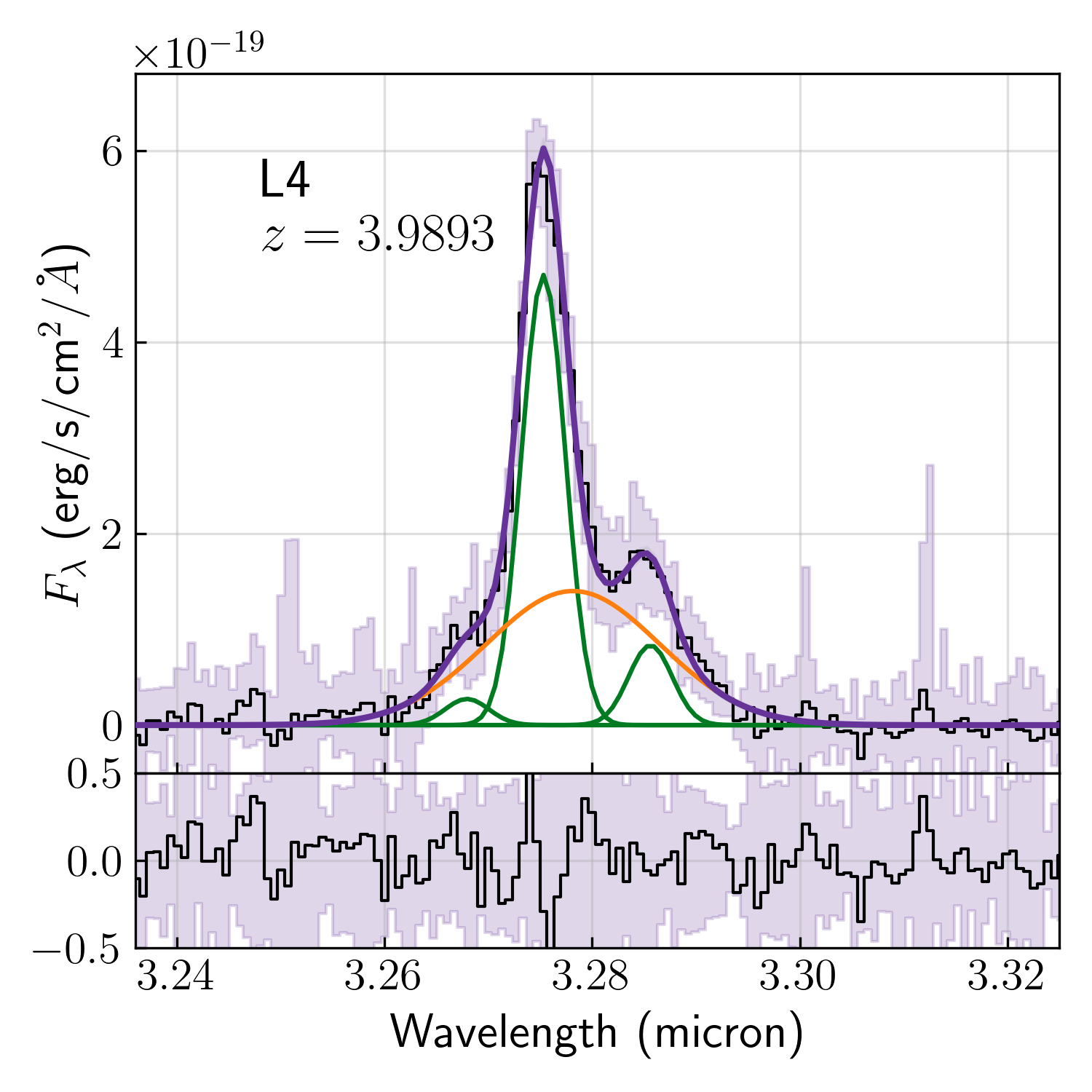}    
    \caption{Best-fitting line profiles for the \ha+\nii\ complex seen across all the identified continuum (C1-C6) and line (L1-L6) emitting regions in the data cube, with per pixel uncertainties shown in purple shaded regions. A large variety of line profiles and kinematics are seen across the system, with certain line emitting regions showing the presence of both broad (orange) and narrow (green) line components, tracing disturbed kinematics and the presence of outflows.}
    \label{fig:linefits}
\end{figure*}

To gain further insights into the complexity of this system traced by gas kinematics, we perform model fitting to the \ha+\nii\ emission, which is the brightest feature in the data cube. The line fitting was performed on the continuum-subtracted spectra, where the continuum was locally measured from line-free regions on either side of the emission lines. The best-fitting models for the emission lines were obtained as follows. We use \texttt{lmfit} to fit model Gaussian functions to the observed data. Following \citet{saxena24}, we allow the fitting of two Gaussian functions to the \ha\ line and a single Gaussian function to each of the two weaker \nii\ lines. Allowing for a broadened \ha\ component enables us to investigate whether there is evidence for (1) a broad-line region (BLR;  traced by broad \ha\ but absence of broad \nii), or (2) out-flowing or turbulent gas in the narrow line region (NLR; traced by broad emission of both \ha\ and other forbidden lines). The width of the narrow \ha\ and \nii\ models are fixed. To obtain the best-fit model, we use chi-squared minimization. Fits using single Gaussian models were performed for the \oi\ and \sii\ lines, owing to their relatively low S/N in the spectra. In Figure \ref{fig:linefits} we show the results of our model fitting to the \ha+\nii\ complex for each region.

C2, the presumed location of the radio AGN host, is best fitted with broad and narrow emission components in \ha. We note, however, that a similarly good fit is also returned when the \nii\ doublet was also allowed to have broad components as an isolated experiment, with the widths of the broad components fixed across \ha\ and \nii. Additionally, L2 and L4 also require a broad \ha\ component to fully match the observed emission line profile of the \ha+\nii\ complex. L3 is well described by a single, broad component for both \ha\ and \nii, implying that a fast outflow or turbulence dominates the gas at this location. We note once again that C2 and L2 appear to straddle the northern radio hotspot, while L3 and L4 appear to straddle the southern radio hotspot. We suggest that turbulence injected in the line-emitting gas by the radio jets may explain why broad kinematical components are preferentially found at these locations.

\subsection{Investigating the presence of broad-line regions}

To further investigate whether the broad components seen in \ha\ may be tracing a dense BLR representing accretion surrounding an active black hole, in this section we explore whether the well-detected \siii\ forbidden line doublet also shows broad components in components C2, L2 and L4. We use \siii\ for this test instead of \sii, as the \sii\ doublet is not spectrally resolved, making it more difficult to observe the presence of any broad component. 

The critical density of the \siii\,$\lambda\lambda9069, 9531$ lines is $\approx 2\times 10^3$\,cm$^{-3}$, which means that for densities $\gg10^3$\,cm$^{-3}$, the \siii\ doublet will be suppressed by collisional de-excitation. Any observed broad components in \siii\ therefore must originate from fast moving low density gas such as that in outflows, and not from a dense BLR around an accreting black hole. The presence of broad features detected in \siii\ can thus be used to rule out that any of the broad features in \ha\ are due to a BLR. 

\begin{figure*}
    \centering
    \includegraphics[width=0.3\linewidth]{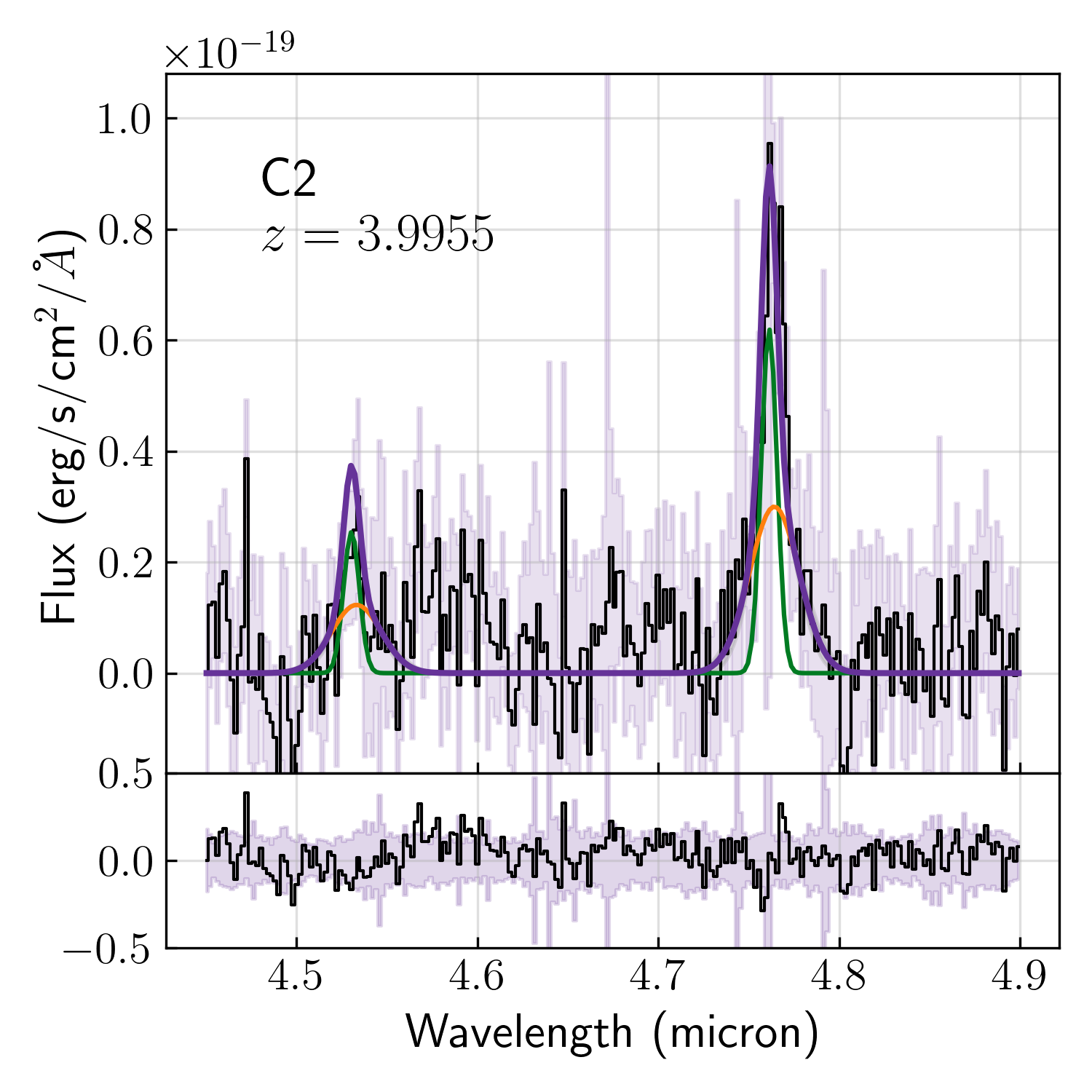}
    \includegraphics[width=0.3\linewidth]{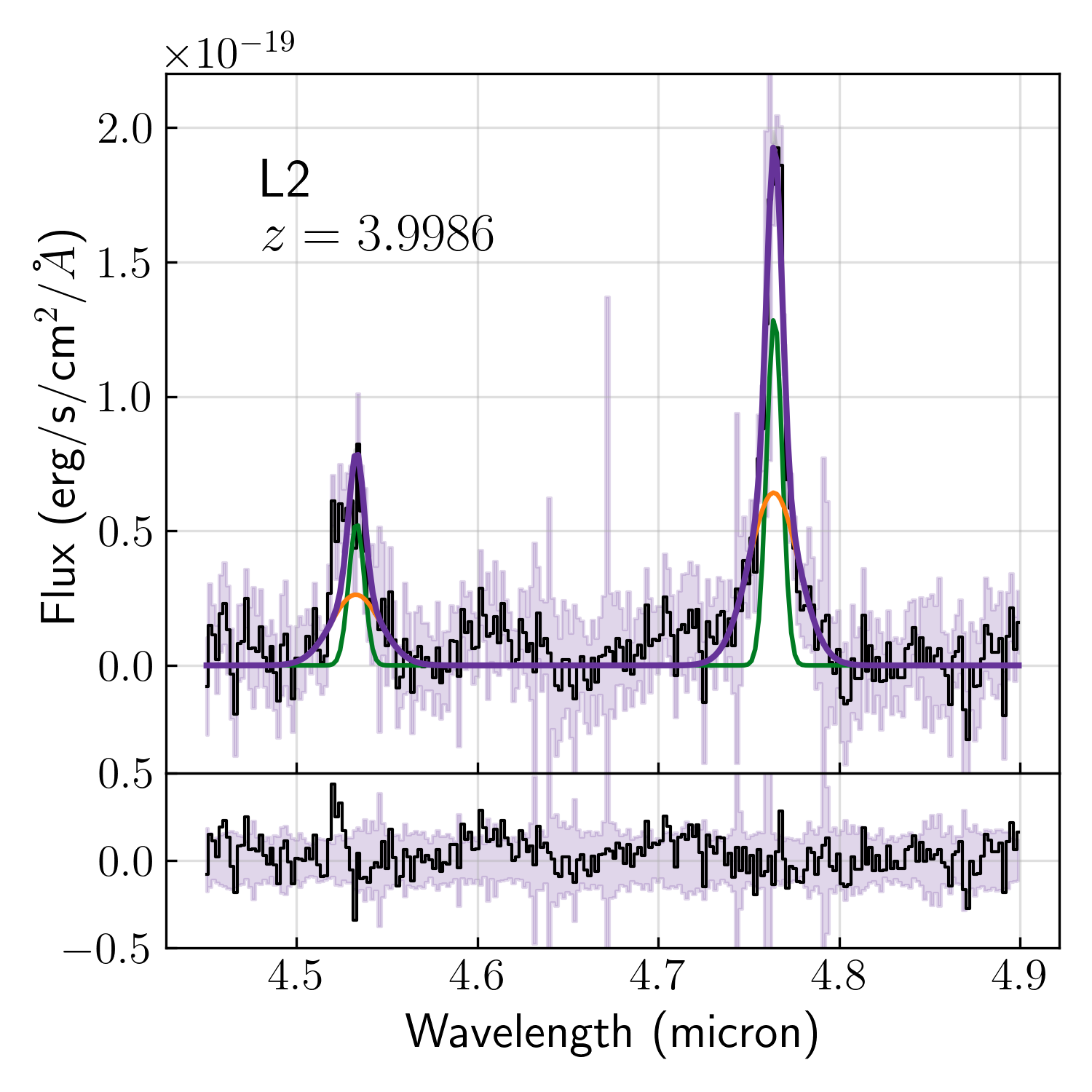}
    \includegraphics[width=0.3\linewidth]{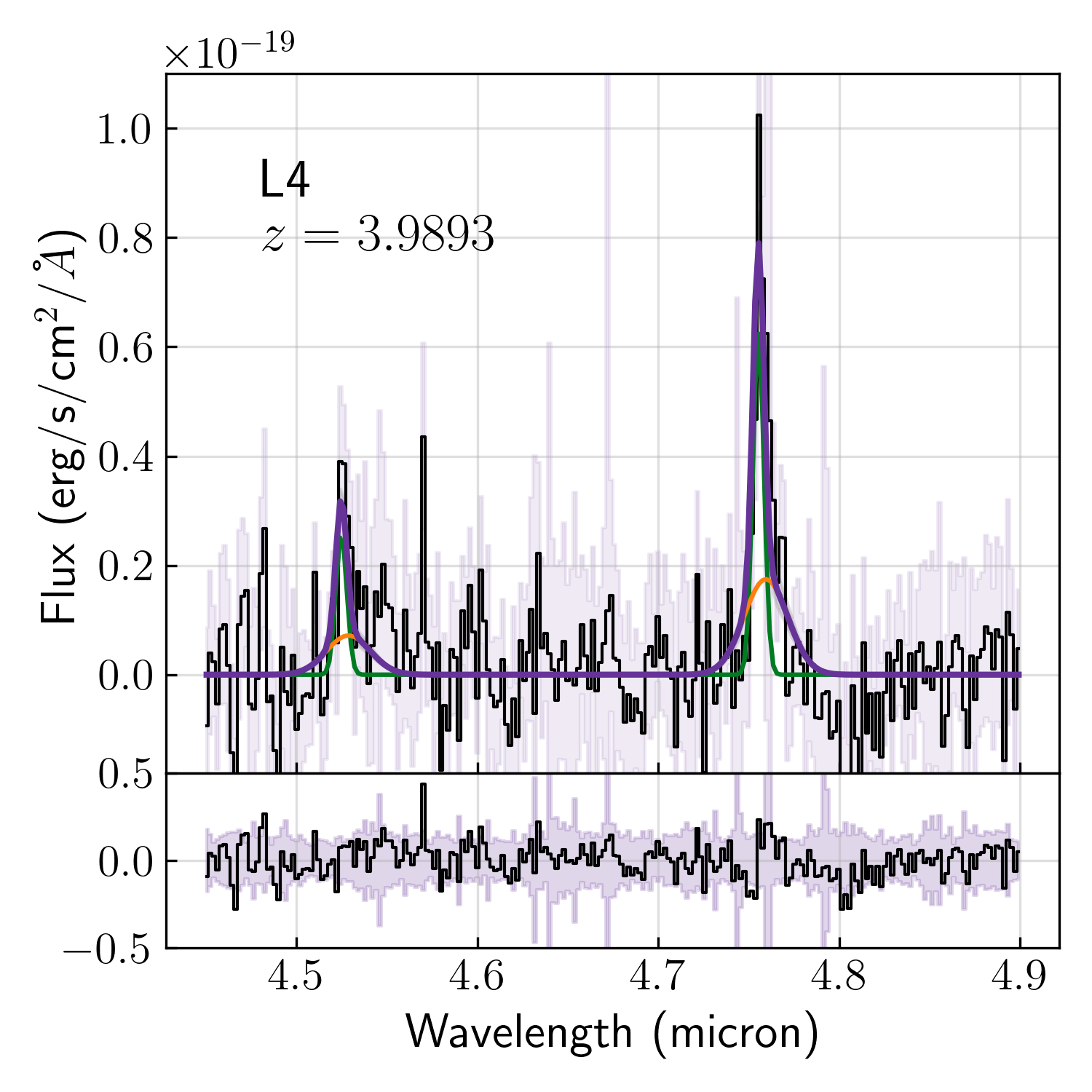}
    \caption{Best-fitting line profiles for the \siii\,$\lambda\lambda9069,9531$ doublet for components C2, L2 and L4 where a broad \ha\ component was identified. By fixing the width of the line component when fitting \siii, we find that the combination of narrow (green) and broad (orange) components is needed to best model the \siii\ doublet emission, thereby confirming that the broad component in \ha\ does not correspond to a BLR, but instead traces outflows.}
    \label{fig:siii_fitting}
\end{figure*}

Using the same methodology as earlier, we perform narrow and broad component fits to the \siii\ doublet emission for components C2, L2 and L4, fixing the line widths (in velocity space) to those measured for \ha, with the results shown in Figure \ref{fig:siii_fitting}. The \siii\ emission seen in C2, L2 and L4 clearly requires the inclusion of a broad component to fit the observed line profile. Since the width of the components was fixed to that of \ha\ emission, this demonstrates that the broad line emission likely does not originate from a BLR surrounding an accreting black hole. In this case, the broadening is likely a result of outflows at these locations. Particularly interesting is C2, the suspected location of the radio-AGN. The broad component has a width of about 2000 km s$^{-1}$ (FWHM), implying a strong galaxy-scale outflow driven by the AGN. This is similar to what was observed for TNJ1338, another $z\approx4$ radio galaxy \citep{saxena24}.

The broad components seen in L2 and L4, and the overall turbulent kinematical structure seen in L3 may be attributed to the interaction between the radio jets and gas clumps near the AGN host galaxy. The northern lobe is likely to deposit energy in a gas cloud at L2, while the southern lobe is likely to drive through L3 and L4, leading to kinetic energy transfer into the gas clouds at these locations. We note that these gas clouds are present between the continuum bright sources, which may have been transported there due to ongoing merger activity.

In Table \ref{tab:line_measurements} we show the results of our line fitting, giving the line fluxes, widths, and ratios measured from fitting narrow and broad components to the \ha+\nii\ emission lines. We additionally show the ratios of low ionization emission lines such as \oi\,$\lambda6300$, \nii\,$\lambda\lambda6548,6583$ and \sii\,$\lambda6716,6731$ to that of \ha, which are typically employed in combination with other emission line ratios to investigate the dominant sources of ionization.

\begin{table*}[t]
    \centering
    \caption{\textup{Emission line fluxes, widths and ratios calculated from emission line fitting for \ha+\nii, \sii\ and \oi\ using the 1D spectra of all continuum and line components with corresponding IDs in the data cube. For components where both narrow and broad components were needed to fit \ha, we report the line fluxes and widths of both components. All line fluxes are given in units of $\times 10^{-18}$\,\flux.}}
    \begin{tabular}{l c c c c c c c c c}
    \toprule 
    ID & $z$ & $F_{\mathrm{H\alpha}}$ & $F_{\mathrm{\nii}6583}$ & $F_{\mathrm{H\alpha}}^{\mathrm{broad}}$ & FWHM & FWHM$^{\mathrm{broad}}$ & \nii/\ha\ & \sii$_{6716,6731}$/\ha\ & \oi$_{6300}$/\ha \\
    & & & & & (\kms) & (\kms) & \\
    \hline
    C1 & $3.9961$ & $27.2 \pm 0.9$  & $15.2 \pm 0.6$ & $-$ & $937 \pm 30$ & $-$ & $0.57 \pm 0.02$ & $0.40 \pm 0.03$ & $0.12 \pm 0.02$ \\
    C2 & $3.9955$ & $10.1 \pm 0.5$ & $0.9 \pm 0.3$ & $53.9 \pm 1.2$ & $236 \pm 12$ & $2026 \pm 5$ & $0.09 \pm 0.02$ & $0.90 \pm 0.07$ & $0.13 \pm 0.02$ \\
    C3 & $3.9848$ & $19.7 \pm 0.6$ & $9.7 \pm 0.5$ & $-$ & $561 \pm 17$ & $-$ & $0.50 \pm 0.02$ & $0.71 \pm 0.05$ & $0.06 \pm 0.01$ \\
    C4 & $3.9839$ & $5.7 \pm 0.3$ & $1.5 \pm 0.2$ & $-$ & $204 \pm 10$ & $-$ & $0.27 \pm 0.03$ & $0.12 \pm 0.02$ & $0.09 \pm 0.01$ \\
    C5 & $3.9883$ & $9.0 \pm 0.3$ & $5.4 \pm 0.3$ & $-$ & $385 \pm 13$ & $-$ & $0.59 \pm 0.02$ & $0.23 \pm 0.02$ & $0.18 \pm 0.03$\\
    C6 & $3.9854$ & $66.2 \pm 0.9$ & $8.6 \pm 0.4$ & $-$ & $448 \pm 6$ & $-$ & $0.13 \pm 0.01$ & $0.46 \pm 0.02$ & $0.24 \pm 0.02$ \\
    \hline
    L1 & $3.9979$ & $38.0 \pm 0.6$ & $22.6 \pm 0.4$ & $-$ & $570 \pm 8$ & $-$ & $0.59 \pm 0.01$ & $0.66 \pm 0.02$ & $0.25 \pm 0.01$\\
    L2 & $3.9986$ & $24.1 \pm 1.6$ & $13.6 \pm 1.0$ & $19.3 \pm 4.6$ & $635 \pm 34$ & $2012 \pm 58$ & $0.56 \pm 0.05$ & $0.64 \pm 0.02$ & $0.31 \pm 0.02$ \\
    L3 & $3.9947$ & $56.7 \pm 0.8$ & $15.0 \pm 0.5$ & $-$ & $2186 \pm 410$ & $-$ & $0.26 \pm 0.01$ & $0.72 \pm 0.10$ & $0.17 \pm 0.03$ \\
    L4 & $3.9893$ & $24.8 \pm 0.5$ & $4.4 \pm 0.8$ & $29.2 \pm 0.5$ & $454 \pm 9$ & $1794 \pm 12$ & $0.18 \pm 0.01$ & $0.49 \pm 0.02$ & $0.21 \pm 0.02$ \\
    \bottomrule
    \end{tabular}
    \label{tab:line_measurements}
\end{table*}

\subsection{Electron density and ionization state of the ISM}

The \sii\ doublet is a powerful diagnostic of the electron density in the ISM, as the \sii\ doublet arises from the metastable energy levels of the $^2D$ state: \sii$\lambda6717: ^2D_{3/2} \rightarrow ^4S_{3/2}$ and \sii$\lambda6731: ^2D_{5/2} \rightarrow ^4S_{3/2}$. In the low-density regime ($n_e \ll 100$\,cm$^{-3}$), collisions are rare and therefore the \sii$\lambda6717$ line is stronger, giving $R = I(\lambda6717)/I(\lambda6731) >1$. As densities increase, increasing collisions cause $I(\lambda6731)$ to become stronger, leading to $R<1$. At the critical density ($n_e \sim 10^3$\,cm$^{-3}$), both lines are equal, giving $R\approx1$. 

To calculate the electron density from the \sii\ doublet ratio, we employ \texttt{pyneb} \citep{pyneb} fixing the gas temperature to $10^4$\,K. Due to a lack of coverage of the \oiii$\lambda4363$ line, which is an accurate temperature diagnostic, it is not possible to estimate a spectroscopically derived electron temperature for the system. We note that a different choice of temperature does not significantly alter the derived electron density from the line ratios \citep{ost06}.

The \sii\ doublet is not clearly spectrally resolved, even at the highest NIRSpec resolution of $R\sim2700$. Therefore, we model the complex as a combination of two Gaussian functions, fixing the width of the Gaussian to match that of the other emission lines seen in the spectrum. The flux ratio of the doublet is then calculated by decomposing the blended \sii\ line into its two components.

We calculate the \siii/\sii\ ratio to place constraints on the ionization state at the locations of the continuum and line-emitting objects in the system. The energy required to ionize S$^{+}$ to S$^{+2}$ is $23.3$\,eV and that required to ionize S to S$^{+}$ is $10.4$\,eV. Therefore, \siii/\sii\ (or S32) ratio gives an indication about the `hardness' of the main ionizing source. Higher S32 indicates more emission arising from the highly ionized regions, making this ratio a powerful diagnostic of the ionization state of the ISM. To calculate the ratio, we sum the \siii\ doublet flux and divide it by the sum of the \sii\ doublet flux. The measured $n_e$ and S32 ratios for all components are given in Table\,\ref{tab:density_ionization}.

The highest $n_e = 1670 \pm 140$ cm$^{-3}$ is found in C3, along with a high S32 ratio of $1.21 \pm 0.12$, which is unsurprising as C3 is spatially coincident with the southern radio lobe of the AGN and is likely the site of active jet-gas interaction where shock ionization elevates both the electron density and the ionization parameter. Interestingly, extremely low electron densities $n_e \leq 50$ cm$^{-3}$ are seen at C4, with S32 ratio of $0.72 \pm 0.18$, likely indicative of fairly diffuse ISM in this continuum emitting clump that lies far away from the radio jet axis. On average we see higher S32 ratios in the line-emitting components, which indicates that harder sources of ionization are present at these locations.

In summary, the components in this complex, interacting system show a diverse range of ISM conditions, pointing towards a mixture of ionizing sources and changes in ISM structure due to competition between star formation, ionization due to AGN or shocks, merger activity and jet-gas interactions.

\begin{table}
    \centering
    \caption{\textup{Summary of the measurements of electron density, $n_e$, from the \sii\ doublet, and \siii/\sii\ ratio, which is a proxy for the ionization parameter, from all continuum and line-emitting components in the system.}}
    \begin{tabular}{l c c c}
    \toprule
    ID & $z$ & \sii\,$n_e$ (cm$^{-3}$) & \siii/\sii\ \\
    \midrule
    C1 & $3.9961$ & $400 \pm 15$ & $0.84 \pm 0.09$ \\
    C2 & $3.9955$ & $796 \pm 70$ & $1.07 \pm 0.09$ \\
    C3 & $3.9848$ & $1670 \pm 140$ & $1.21 \pm 0.12$ \\
    C4 & $3.9839$ & $\leq50$ & $0.72 \pm 0.18$ \\
    C5 & $3.9883$ & $600 \pm 50$ & $0.73 \pm 0.08$ \\
    C6 & $3.9854$ & $500 \pm 24$ & $1.43 \pm 0.10$ \\
    \hline
    L1 & $3.9979$ & $510 \pm 20$ & $1.50 \pm 0.06$ \\
    L2 & $3.9986$ & $375 \pm 15$ & $1.55 \pm 0.06$ \\
    L3 & $3.9947$ & $625 \pm 300$ & $1.61 \pm 0.80$ \\
    L4 & $3.9893$ & $560 \pm 20$ & $0.53 \pm 0.03$ \\
    \bottomrule
    \end{tabular}
    \label{tab:density_ionization}
\end{table}

\subsection{ISM-radio jet interactions and gas kinematics}
\label{sec:kinematics}

\begin{figure*}
    \centering
    \includegraphics[width=\linewidth]{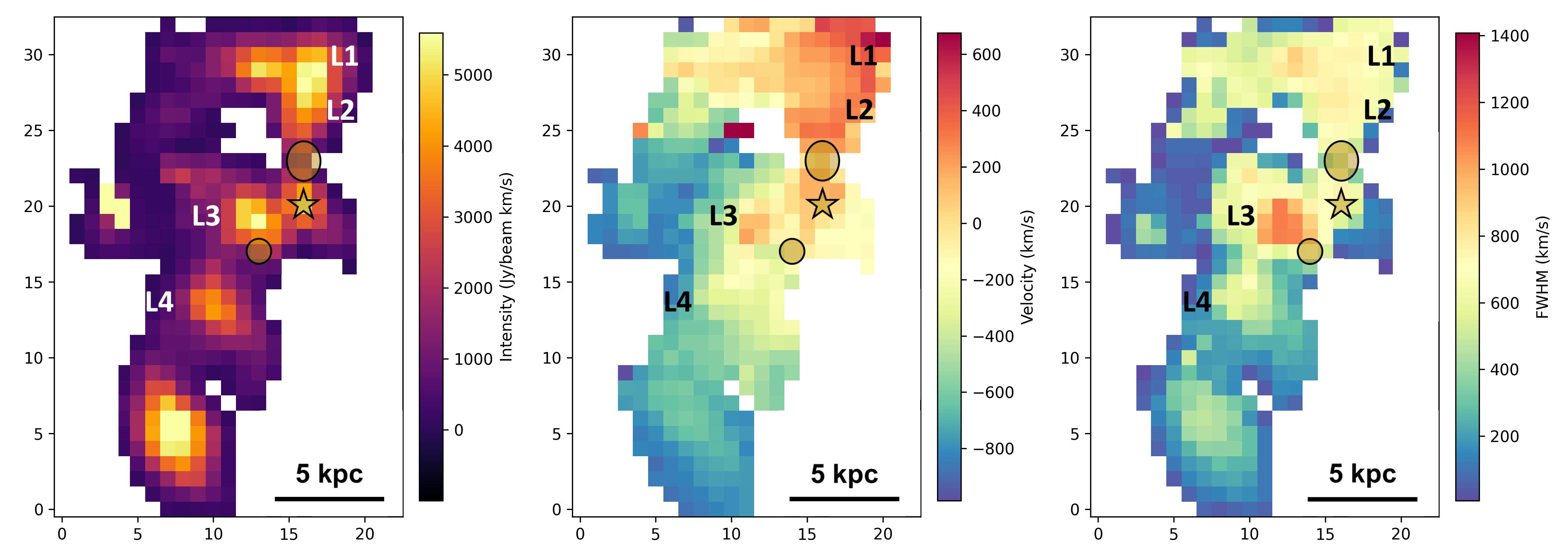}
    \caption{Intensity (left), intensity-weighted velocity or Moment 1 (middle), and FWHM (right) maps centered around the \ha+\nii\ emission from the radio-AGN host C2 in TGSSJ1530. C2 has been marked using a star, and the location of the radio hotspots have been marked using circles. The moment maps reveal complex kinematical structures in this system, with the velocity map suggestive of large scale rotation around the AGN host, indicative of ongoing merging. The FWHM map clearly identifies the location where the southern radio lobe interacts with the gas clumps, which leads to extremely turbulent emission line structures at L3.}
    \label{fig:moments}
\end{figure*}

Three out of four line-emitting components (L2, L3 and L4) require substantial broad line components. Possible sources of this disturbed gas are the passage of the radio jet, a radiative wind from the obscured AGN, or the merging activity. To further explore this, in Figure\,\ref{fig:moments} we show maps of the \ha\ intensity (left), intensity-weighted velocity or Moment 1 (middle) and the FWHM or Moment 2 (right). The velocities are shown with respect to the velocity centroid of the \ha\ emission of component C2. In the Figure, the presumed location of the radio-AGN host (component C2) has been highlighted with a star, and the location of the two radio hotspots either side of the AGN have been highlighted using circles.

Line-emitting region L3 is the closest component to the southern radio lobe, and shows high intensity and the broadest lines, as was previously demonstrated using 1D line fitting. Therefore, L3 likely corresponds to the location where the radio jet efficiently couples to the ISM, exciting it and introducing significant turbulence. The lack of clear continuum emission at this location suggests that there are no stars in this region, consistent with the interpretation of jet-gas interactions. The line emitting region L2 that appears closest to the northern lobe shows evidence of relatively minor turbulence (FWHM \,$\approx635$\,\kms) compared to L3, and has less than half the total \ha\ luminosity of L3.

Besides the turbulent region near L3, there are no other distinct velocity features in the map. However, the Moment 1 map shows evidence for global rotation in the complex system, with the northern part significantly red-shifted and the southern part significantly blue-shifted compared to the velocity of the radio AGN. The large scale structure of this rotation rules out simple kinematics arising from a single galaxy. The north-west component (L1) in the system shows high line brightness as well as a disturbed velocity structure, which is co-spatial with a merging companion as seen previously in the continuum images. This structure is qualitatively similar to that observed by \citet{vanojik96}, who concluded that slowly rotating gas extending beyond the radio emission points to gas being accreted from the larger environment. Combined with the evidence for multiple distinct velocity components, the TGSSJ1530 complex appears to be in a state of interaction or merger. 

From this analysis it is clear that although there is some evidence for interactions between the gas and the radio jets, the relatively small physical extent of the radio jet compared to the full linear size of the system suggests that it is of influence only in the inner regions. Although we cannot completely rule out that there is more diffuse or aged radio emission extending further out that is being missed by our current VLBI observations, the large scale gas kinematics is likely dominated by the gravitational interactions between individual galaxies. In Section\,\ref{sec:massive} we use spectro-photometric modeling to calculate stellar masses and ages of the continuum components. This will indicate that there is a substantial number of massive galaxies involved in this merger. 

\section{A forming massive galaxy at $z=4$}
\label{sec:massive}

\subsection{Insights from SED fitting}
\label{sec:seds}

Since both IFU spectroscopy and imaging data are available, the SED fitting is performed in one step on the joint spectro-photometric data set. From the resulting best-fit SEDs we infer key physical properties that shed light on (1) the assembly history and (2) the possible evolutionary state of this complex system. 

The photometry for all continuum components is first extracted from \emph{HST} and \emph{JWST} images by placing circular apertures at the locations of the extracted 1D spectra, with aperture radii of $0\farcs2$ to match the spectroscopic apertures. We note here that the images were resampled to match the pixel scale of the F210M image, representing the closest match to central wavelength of the available photometry. As there is considerable diffuse emission surrounding the continuum sources, we do not perform background subtraction using an annulus surrounding the aperture. Instead we place similarly sized apertures in regions that contain no apparent source flux, and then subtract the median flux in these background apertures from the source fluxes. 

The agreement in fluxes between the extracted 1D spectra and photometric data was validated by convolving the spectra with the NIRCam filters and comparing these to the observed NIRCam fluxes. Excellent agreement was found between the two measurements, which implies that there are no significant zero-point errors at least in the NIRCam images. 

As was shown in Figure \ref{fig:images}, most of the components are `HST-dark', i.e., they are faint at rest-frame UV and bright at redder wavelengths. These objects are similar to the sample of high redshift galaxies presented in \citet{Barrufet2023}. These intrinsically red objects trace massive, dust-obscured galaxies that are typically missed by censuses based on rest-UV selection alone.

The publicly available fitting code \texttt{BAGPIPES} \citep{Carnall2018, Carnall2019} was used to perform spectro-photometric SED fitting for the continuum sources C1-C6. To perform the fitting, we used the default stellar population models that are the 2016 update of the \citet{bc03} models, as described in \citet{Chevallard2016}, which include the \texttt{MILES} spectral library \citep{miles11}. These are single star models that do not include any binary interactions.

We assumed a two-component star-formation history (SFH), which includes an exponentially declining SFH ($\tau$-model) with an additional burst component. For the exponentially declining SFH component, uniform priors were set on the age ranging from $0.3$\,Gyr to the age of the Universe at the source redshift, $\tau$ in the range $[0.3, 15]$, stellar mass formed in the range $[10^8, 10^{12}]$\,$M_\odot$ and stellar metallicity in the range $[0, 1.5$]\,$Z_\odot$, where $Z_\odot$ is the solar metallicity with the value $0.02$.

For the burst component, uniform priors were set on the age in the range $[0.001, 0.020]$\,Gyr, mass formed in the range $[10^6, 10^{10}]$\,$M_\odot$ and metallicity in the range $[0, 1.5$]\,$Z_\odot$. The choice of the priors requires the bulk of the stellar mass to have formed following an exponentially declining history, with the burst component in place to account for any recent star-formation that may be powering the observed emission lines.

We additionally included dust attenuation and nebular emission. We employed the \citet{Calzetti2000} dust attenuation law, with uniform priors on the $A_V$ parameter in the range $[0, 2.5]$. To model the nebular emission, we employed uniform priors on the dimensionless ionization parameter, $\log(U)$, in the range $[-3.5, -2]$. The spectral resolution of the emission lines was matched to the resolution of NIRSpec/IFU spectra.

Fixing the redshifts to the spectroscopically derived redshifts for all the continuum components, SED fitting was performed simultaneously on the photometric and spectroscopic data points. We note here that we did not employ any AGN models in \texttt{BAGPIPES}. This means that the best-fit SED for C2, the suspected host of an (obscured) AGN, may not sufficiently describe its composite spectral components. However, making the reasonable assumption that AGN continuum light does not dominate the rest-frame optical emission of C2 (this is justified by the lack of any BLR emission underlying the Balmer lines), then the stellar-light only models are expected to still provide a reliable estimate of the stellar population parameters.

\begin{table*}
    \centering
    \caption{\textup{Physical parameters (and $1\sigma$ uncertainties) inferred from best-fitting SED models for continuum components in TGSSJ1530, and (SED-derived) dust-corrected star-formation rate from \ha\ using the \citet{Kennicutt1994} relation.}}
    \begin{tabular}{l c c c c c c}
    \toprule
    Component & $z$ & $\log(M_\star/M_\odot)$ & Age (Gyr) & $A_V$ (mag) & $\log(U)$ & SFR$^{\rm{dustcorr}}_{\rm{H}\alpha}$ (\sfr) \\
    \midrule
    C1 & $3.9961$ & $10.34^{+0.02}_{-0.01}$ & $0.44^{+0.03}_{-0.03}$ & $0.95^{+0.06}_{-0.06}$ & $-2.9^{+0.0}_{-0.0}$ & $71.2 \pm 2.3$ \\ 
    C2 & $3.9955$ & $10.64^{+0.02}_{-0.01}$ & $0.75^{+0.13}_{-0.09}$ & $1.84^{+0.06}_{-0.06}$ & $-3.4^{+0.1}_{-0.1}$ & $87.8 \pm 2.5$ \\ 
    C3 & $3.9848$ & $10.68^{+0.01}_{-0.02}$ & $0.92^{+0.04}_{-0.06}$ & $2.48^{+0.01}_{-0.03}$ & $-3.3^{+0.1}_{-0.1}$ & $163.0 \pm 5.7$ \\ 
    C4 & $3.9839$ & $9.45^{+0.02}_{-0.03}$ & $0.61^{+0.11}_{-0.08}$ & $0.04^{+0.06}_{-0.03}$ & $-3.3^{+0.2}_{-0.1}$ & $8.3 \pm 0.4$ \\ 
    C5 & $3.9883$ & $10.56^{+0.01}_{-0.02}$ & $1.17^{+0.03}_{-0.04}$ & $2.41^{+0.06}_{-0.11}$ & $-3.4^{+0.1}_{-0.1}$ & $70.2 \pm 2.5$ \\ 
    C6 & $3.9854$ & $8.8^{+0.05}_{-0.04}$ & $0.17^{+0.06}_{-0.03}$ & $0.62^{+0.03}_{-0.03}$ & $-3.2^{+0.1}_{-0.1}$ & $134.7 \pm 1.8$ \\ 
    \bottomrule
    \end{tabular}
    \label{tab:sed_fitting}
\end{table*}

\begin{figure}
    \centering
    \includegraphics[width=\linewidth]{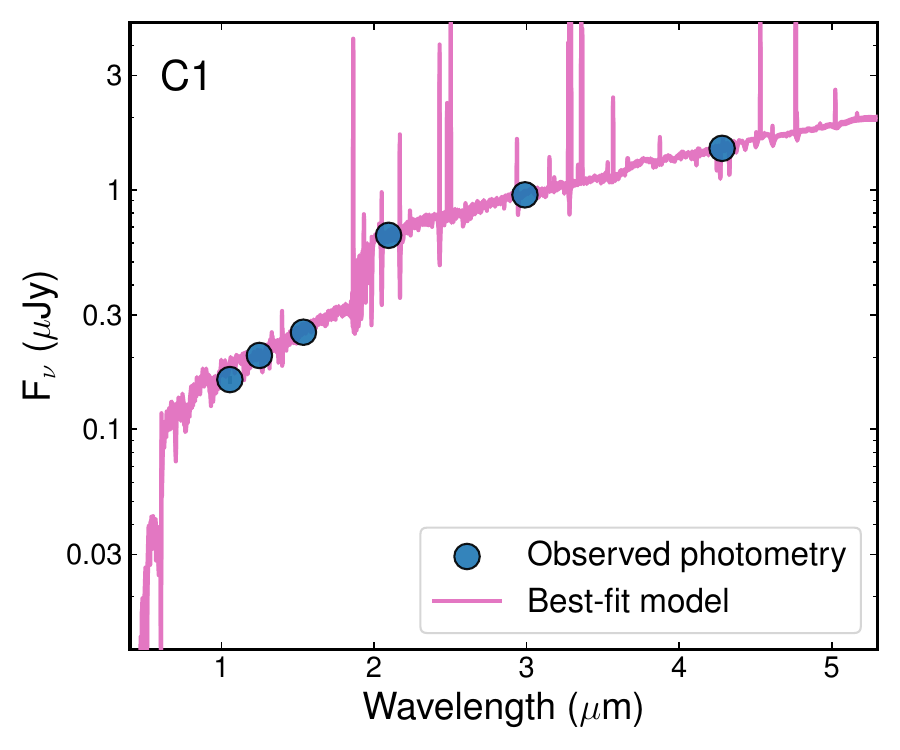}

    \includegraphics[width=\linewidth]{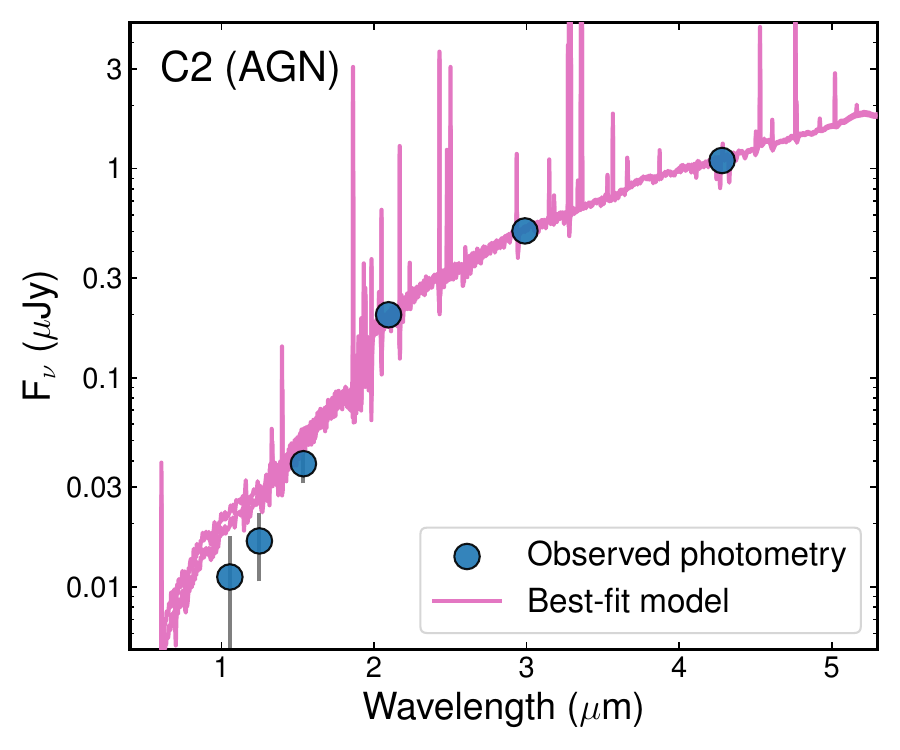}
    \caption{Best-fitting model SEDs derived for component C1 (top) and C2 (bottom), which is also the likely host of the radio AGN in this system. The best-fitting SEDs are able to reproduce the strong Balmer break seen from photometry, and the emission lines seen from extracted 1D IFU spectra. The presence of the Balmer breaks clearly indicates the presence of an older stellar population, pointing towards a relatively early formation timescale for these two galaxies that are at the heart of the merger in this system.}
    \label{fig:sed_fits}
\end{figure}

The key physical properties measured for each continuum-emitting component from SED fitting are summarized in Table~\ref{tab:sed_fitting} and the best-fitting SEDs for components C1 and C2 for demonstration purposes are shown in Figure\,\ref{fig:sed_fits}. Overall, the TGSSJ1530 system contains several high stellar mass and dusty components. Besides C2 with a stellar mass of $\log(M_\star/M_\odot) = 10.64$, the suspected AGN host galaxy, three more components have similarly high stellar mass ($\log(M_\star/M_\odot) = 10.3-10.7$). We also point out that C2 is the only continuum source evaluated that lies within the radio continuum structures (see Fig. \ref{fig:composite_radio}). C2, C3 and C5 are the most massive, oldest and dustiest components and are part of a crowded region, where perhaps dust has accumulated. An arc- or bridge-like feature connects C5 to C4. The latter is relatively young and has negligible dust. Our interpretation is that the entire complex is involved in interactions and mergers. C6, which is located furthest away from the radio AGN, is consistent with a low-mass, actively star-forming galaxy. 

\subsection{Star-formation in the system}

In addition to the inferred physical parameters from SED fitting discussed in the previous sub-section, we also measured the current rate of star-formation from the \ha\ emission in all continuum-emitting components in this system, accounting for the dust attenuation. For C2 where both broad and narrow \ha\ emission components are clearly identified, we can attribute (at least) the narrow component to be originating from young star formation. On the other hand, we caution that an estimate of SFR for this galaxy is likely to be an upper limit, owing to the possible presence of non-thermal ionization sources in C2 as has been noted previously. 

To calculate SFRs from (narrow) \ha\ emission, we first applied dust-correction using the $A_V$ value derived from best-fitting SEDs employing the \citet{Calzetti2000} dust attenuation law. We then used the conversion factor derived by \citet{Kennicutt1994} using single-star models and a Salpeter IMF with an upper-mass cutoff of 150\,$M_\odot$, consistent with models used in the SED fitting. The intrinsic SFR was then derived from the dust-corrected \ha\ luminosity, $L_{\rm{H}\alpha}$ using the relation $\rm{SFR} = L_{\rm{H}\alpha}/1.26\times10^{41}$\,$M_\odot$\,yr$^{-1}$. The dust-corrected SFRs calculated from \ha\ are also given in Table \ref{tab:sed_fitting}. 

\begin{figure}
    \centering
    \includegraphics[width=\linewidth]{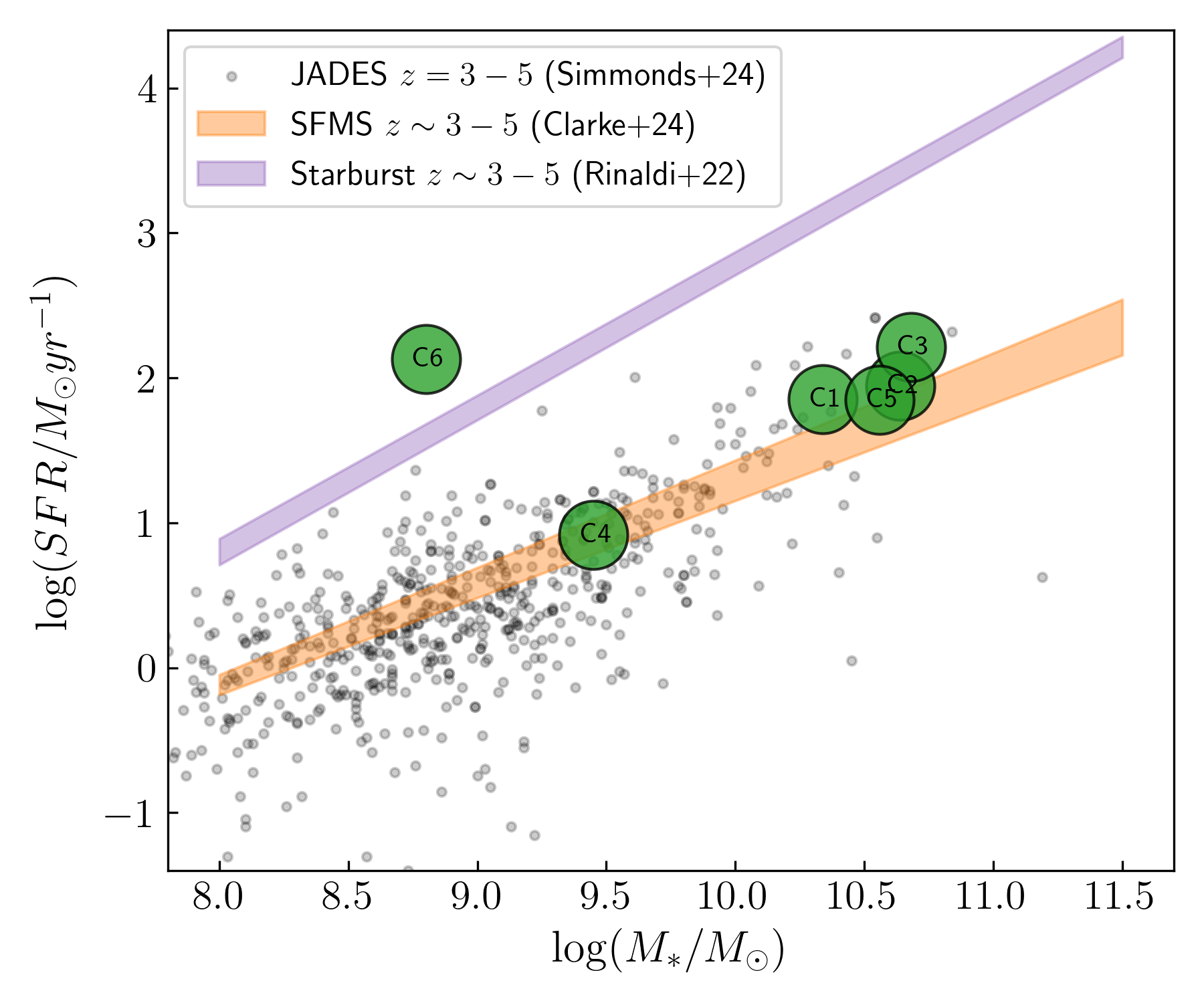}
    \caption{The star-forming main sequence (SFMS) showing the locations of the six continuum-detected galaxies (with the radio AGN being traced by C2) at $z\sim4$. Also shown for comparison is the SFMS derived from normal star-forming galaxies using \ha\ emission in the same redshift range by \citet{Clarke2024}, and the sequence for starburst galaxies at similar redshifts derived by \citet{Rinaldi2022}. Individual data points trace SFR and stellar mass measurements from broadband photometric SED fitting of galaxies in JADES \citep{Simmonds2024}. Overall, we find that galaxies in the TGSSJ1530 system lie slightly above or close to the SFMS, with C6 clearly tracing a starburst galaxy in the system, which is also furthest away from the merging complex.}
    \label{fig:sfms}
\end{figure}

The large majority of galaxies in this system appear to be forming stars at a relatively high rate. The highest SFR is seen in C3, with 163\,\sfr, followed by C6 with $135$\,\sfr. C1, C2 and C5 are forming stars with rates between $70-90$\,\sfr, with C4 exhibiting the lowest SFR of $8$\,\sfr. The total dust-corrected SFR in the system is $\approx555$\,\sfr. 

To put the SFRs measured across all 6 galaxies in this system into context, in Figure\,\ref{fig:sfms} we show the distribution of dust-corrected SFRs and stellar masses with respect to the star-forming main sequence (SFMS) derived for normal star-forming galaxies at $z\sim4$ from \citet{Clarke2024} using \ha-data, and for starburst galaxies by \citet{Rinaldi2022} using rest-frame UV data (see also \citealt{Rinaldi2025}). We additionally show the SED-fitting derived SFRs for galaxies in the JADES survey by \citet{Simmonds2024}.

With the exception of C6, all galaxies seem to lie very close to the SFMS, with C1 and C3 lying marginally above the sequence but within the observed scatter. C6 on the other hand lies much closer to the sequence measured for starburst galaxies, lying clearly above the best-fitting relation but once again within the individual scatter determined by \citet{Rinaldi2022}. Given that C6 is an extended source, located the furthest away from the radio AGN and also has the lowest inferred stellar mass and mass-weighted age, the high SFR confirms this object as a distinct, companion starburst galaxy.

\subsection{Comparison with other systems}

\subsubsection{Other HzRG fields observed with NIRSpec}

Besides TGSSJ1530, five other HzRGs at $z\sim4$ have been observed with the NIRSpec/IFU, providing a convenient sample for comparison. TN J1338$-$1942 is a large, elongated continuum source with diffuse and clumpy emission line gas that lies almost entirely within the radio structure \citep{duncan23,saxena24}. The source shows evidence for strong jet-gas interactions. No distinct companion galaxies were identified within the NIRSpec field.

\citet{wang25a} studied the direct environments of four more HzRGs at $z\approx3.5$ \citep[4C$+$03.24, 4C$+$19.17, TN J0121$+$1320, TN J0205$-$2242; see also][]{wang24,wang25b}. A search for close companions identified by \oiii\ (NIRSpec) or \almacii\ line emission (ALMA) resulted in one \oiii\ emitter and one to several \almacii\ companions per HzRG field. Some of these could be companion galaxies merging with the HzRG, while other could be in-falling ionized or cold gas clouds or distinct AGN wind features. 

The most striking difference between the TGSSJ1530 field and the five comparison fields referenced above are the multiple, distinct companions found around TGSSJ1530. The six (continuum) objects that form the TGSSJ1530 system thus represent a significant excess. Furthermore, there are at least three objects in the TGSSJ1530 system that are of comparable stellar mass to the object hosting the radio source, whereas the companions studied by \citet{wang25a} appear much less massive than their radio galaxy.   

\subsubsection{Dense galaxy groupings at high redshift}
SDSS J165202.64+172852.3 is a quasar at $z=2.9$ with several companions within projected distances of 15 kpc \citep{wylezalek22}. At least three of the companions are high stellar mass objects ($\gtrsim10^{10}$ $M_\odot$), while the quasar host galaxy is at least an order of magnitude more massive. Of the comparisons presented here, this particular system is probably the one that most closely resembles the environment of TGSSJ1530.   

The central region of the protocluster core SPT2349-56 at $z=4.3$ shows three sources detected at 850 $\mu$m and [CII] within an area roughly similar to the NIRSpec/IFU \citep{chapman24}. Only one of these is detected in the rest optical in an HST F160W image. MeerKAT 816 MHz and ATCA 2.2 GHz data have shown that this optical source (``C") likely hosts a radio AGN with a radio luminosity at a rest-frame frequency of 1.4 GHz about ten times lower than that of TGSSJ1530 \citep{chapman24}. SPT2349-56 and TGSSJ1530 are thus at least qualitatively similar in the fact that both are identified with a radio AGN in a dense environment. 

CGG-z5 is a group of six galaxies having photometric redshifts consistent with $z\approx5.2$ that lie within a $1.5\arcsec\times3\arcsec$\ region \citep{jin23}. The group is dominated by a $10^{10}$ $M_\odot$ stellar mass galaxy surrounded by less massive companions. Although the general configuration appears similar to that seen in the TGSSJ1530 field, the total stellar mass scale of CGG-z5 appears lower by at least a factor of 10. 

Lastly, SPT0311--58 consists of a lensed pair of dusty star-forming galaxies at $z\approx6.9$. The pair is separated by about 8 kpc and has a total SFR of $>3000$ $M_\odot$ yr$^{-1}$. \citet{arribas24} recently used the NIRSpec/IFU to find a remarkably clumpy region of optical line and continuum sources. The number of clumps in SPT0311--58 is substantially larger than that in TGSSJ1530, but the majority has much lower stellar masses ($M_\star<9.5$ M$_\odot$) and a larger range of velocities (1500 km s$^{-1}$).   

\subsection{Halo mass of the overdensity}

In the previous section we have shown that the TGSSJ1530 field hosts a rare concentration of massive galaxies. The total stellar mass in components C1-C6 is $\approx1.5\times10^{11}\,M_\odot$. We note that this mass is unchanged if we discard the two lowest mass sources (C4 and C6), and we also did not include any contribution from the total ongoing SFR of order 500\,\sfr. 

From \citet{behroozi19} we find the typical number density of such sources at $z\sim4$ of $\sim3\times10^{-5}$ Mpc$^{-3}$ dex$^{-1}$ (assuming a single source of this mass). The typical stellar mass to dark matter halo mass ratio is in the range 0.005--0.01. This implies a total dark matter halo mass of at least $10^{13}$ $M_\odot$. Such halos are very rare at these redshifts, and correspond to a peak height of 4.5 and halo bias of 9.8 \citep{diemer18}. The radius $R_{200}$ of such a halo would be about 133 kpc, much larger than full extent of the system seen in the IFU image (full diameter of about 21 proper kpc). 

It is difficult to derive exactly what would be the future state of this system, but two basic conclusions seem unavoidable. First, the four massive objects are likely to merge within a few Gyr, given their small (projected) separations and relatively small velocity differences \citep[see][for example]{shi24}. Second, halos as massive as $10^{13}$ $M_\odot$ at $z\sim4$ typically evolve into $10^{14-15}$ $M_\odot$ systems at $z=0$ \citep{chiang13}. However, we must emphasize that this conclusion is entirely based on a stellar mass argument. The fact that we find at least four massive, highly clustered galaxies is consistent with such a proto-cluster interpretation, but at present there is no evidence to suggest that this object is part of a large overdensity of galaxies on Mpc scales, as may be expected if this truly is a massive, forming structure. 

It is also important to point out that there are far more massive systems that can be found at $z\sim4$ \citep{remus23}, although these are not representative for typical massive systems today. Such systems likely form through a sequence of rapid mergers at high redshift. It seems unlikely that the C1-C6 components have reached an equilibrium at the centre of a single massive halo. 

Observing such a number of fairly massive galaxies at a projected separation of about 10 kpc suggests we may be catching TGSSJ1530 at a special time, qualitatively similar to the few per cent of BCGs that are found to assemble early in simulations \citep{rennehan20}.

\section{Conclusions}
\label{sec:conclusions}

In this paper, we have presented new \emph{JWST} NIRSpec IFU and NIRCam observations for TGSSJ1530+1049 (``TGSSJ1530''), which was previously identified erroneously as a radio-loud AGN at $z=5.72$. The new observations unambiguously confirm that the system is a radio galaxy at $z=4.0$. The NIRSpec IFU and NIRCam imaging reveal a highly complex, likely merging system of at least six continuum-bright objects surrounding the radio source. One object, in particular, is marked as the likely host galaxy of the radio AGN due to its position in between the radio hotspots.

We highlight the presence of at least six continuum sources and an additional four predominantly line-emitting sources, amidst fainter and more diffuse features. It remains unclear whether the presence of the radio AGN is purely coincidental, or whether the AGN was triggered by gas flows in the complex merger environment. Strong alignment of the \ha\ emission line morphology (but not the continuum) is seen along the radio axis, suggestive of a biconical zone of illumination by a central obscured AGN. However, the observation that some of the brightest line emission clumps appear (1) well beyond the radio structure, (2) have significant velocity offsets and (3) strong underlying continuum consistent with stellar populations, suggests that this alignment is, at least partly, caused by the particular configuration of interacting or infalling galaxies.    

Although broad components are found in the \ha\ emission of the likely host galaxy of the radio AGN and several other components, the presence of BLR emission from AGN accretion disks can be ruled out on the basis of finding similarly broad components in forbidden line transitions. Therefore, the broadening of the emission lines is most likely tracing galaxy-scale outflows driven by the AGN or star formation.

As the line emission extends well beyond the location of the radio hotspots as inferred from VLBI data, the emission line gas further out is assumed to be unaffected by the current radio jet activity. In some regions, however, kinematical modeling of the emission lines shows evidence for jet-gas interaction, in particular a region around the Southern radio hotspot broadened to $>1000$ km s$^{-1}$ (FWHM). Furthermore, moment maps of the \ha\ emission show a large-scale velocity gradient across the full extent of the IFU field (projected diameter of about 21 physical kpc at $z=4$), which is suggestive of a rotating system.  

Leveraging \emph{HST} and \emph{JWST} imaging and performing spectro-photometric SED fitting, the stellar masses and SFRs of all continuum-bright sources in the TGSSJ1530 field were constrained. The stellar masses of the six main continuum components range from $\log(M_\star/M_\odot) = 8.8 - 10.7$, with the radio AGN having a total stellar mass of $\log(M_\star/M_\odot) = 10.6$. The (dust-corrected) star-formation rates range from $8-163$\,\sfr, with the radio AGN having a star-formation rate of $88$\,\sfr. Overall, five out of the six continuum-bright galaxies lie close to the SFMS at $z\sim4$, with the lowest mass companion being more akin to a typical starburst galaxy. 

With at least four objects having a stellar mass in the range $10.3<\mathrm{log}(M_\star/M_\odot)<10.7$ within a $10\times10$ (projected) kpc$^2$ area, TGSSJ1530 is one of the densest regions found at these redshifts to date. The number of companions is much larger than that observed in five other HzRG fields observed with NIRSpec \citep{saxena24,wang25a}, but is quite comparable to the dense environment encountered around a high redshift quasar \citep{wylezalek22}. The system also shows some general similarities with other dense structures and environments \citep{chapman24,arribas24,jin23}. Using a simple stellar mass-based analysis, we predict that the TGSSJ1530 system corresponds to a total dark matter halo mass of at least $\approx 10^{13}\,M_\odot$. Based on the physical separations and velocity differences between the galaxies, it is likely that all of these galaxies will merge to form a massive galaxy within a few Gyr. Although there currently is no evidence for a wider protocluster structure, a halo of this size at $z\sim4$ would likely evolve into a $10^{14}-10^{15}\,M_\odot$ system by $z\sim0$. 

TGSSJ1530 offers yet another rare view on one of the early stages of the formation of a BCG-type galaxy in the local Universe. The TGSSJ1530 system is consistent with a subset of BCGs in numerical cosmological simulations that accumulated a large fraction of their mass early in a rapid series of mergers at high redshift \citep{rennehan20,remus23}. Future work could further quantify its large-scale overdensity, which could result in more detailed estimates of the total mass and the future fate of the system. This work has shown that the identification of candidate HzRGs from purely radio-selected samples continues to deliver interesting probes of cosmology, massive galaxy formation and supermassive black holes. 

\section*{acknowledgments}
We thank the referee for a constructive report that improved the quality of this manuscript. AS and RAO thank Mingyu Li for raising the possibility of the initially reported rest-frame UV line to possibly be He\,\textsc{ii} emission. AS acknowledges funding from the “FirstGalaxies” Advanced Grant from the European Research Council (ERC) under the European Union’s Horizon 2020 research and innovation programme (Grant agreement No. 789056). RO was supported by a productivity grant (302981/2019-5) from the National Council for Scientific and Technological Development (CNPq). MVM acknowledges support by grant Nr. PID2021-124665NB-I00 by the Spanish Ministry of Science and Innovation/State Agency of Research MCIN/AEI/10.13039/501100011033 and by ``ERDF A way of making Europe''. KJD acknowledges support from the STFC through an Ernest Rutherford Fellowship (grant number ST/W003120/1). K\'EG and SF was supported by the Hungarian National Research, Development and Innovation Office (NKFIH excellence grant TKP2021-NKTA-64). LP acknowledges support from INAF Large grant
2022 ``Extragalactic Surveys with JWST'' and  PRIN 2022 MUR project 2022CB3PJ3—First Light And Galaxy Assembly (FLAGS) funded by the European Union—Next Generation EU.

This work is based on observations made with the NASA/ESA/CSA James Webb Space Telescope. The data were obtained from the Mikulski Archive for Space Telescopes at the Space Telescope Science Institute, which is operated by the Association of Universities for Research in Astronomy, Inc., under NASA contract NAS 5-03127 for JWST. These observations are associated with program GO 1964. This research is also based on observations made with the NASA/ESA Hubble Space Telescope obtained from the Space Telescope Science Institute, which is operated by the Association of Universities for Research in Astronomy, Inc., under NASA contract NAS 5–26555. These observations are associated with program GO 16693.

This research would not have been possible without the thousands of hours of effort put in by the open-source development community around the world.


\bibliography{hzrgs}{}
\bibliographystyle{aasjournal}

\end{document}